\documentclass{IEEEtran}
\def\BibTeX{{\rm B\kern-.05em{\sc i\kern-.025em b}\kern-.08em
    T\kern-.1667em\lower.7ex\hbox{E}\kern-.125emX}}

\usepackage{amsmath,amssymb,amsfonts}
\usepackage{mathabx}
\usepackage{algorithmic}
\usepackage{graphicx}
\usepackage{textcomp}
\usepackage[utf8]{inputenc}
\usepackage[T1]{fontenc}
\usepackage{amsmath}
\usepackage{amsfonts}
\usepackage{amssymb}
\usepackage{balance}
\usepackage{booktabs}
\usepackage{multirow}
\usepackage{graphicx}
\usepackage[rightcaption]{sidecap}
\usepackage{import}
\usepackage{textcomp}
\usepackage[dvipsnames]{xcolor}
\usepackage{bbm}
\usepackage{slashed}
\usepackage{caption}
\usepackage{subcaption}
\usepackage{lipsum}
\usepackage{enumitem}
\usepackage{tabularx}
\usepackage{xcolor}
\usepackage[ruled,linesnumbered]{algorithm2e}
\usepackage{leftindex}
\usepackage{mathtools}
\usepackage[noadjust]{cite}
\usepackage{wrapfig}
\usepackage{hyperref}
\hypersetup{hidelinks=true}

\usepackage{easyReview}
\usepackage{bm}

\SetKwComment{Comment}{/* }{ */}
\SetKwInOut{Input}{Input}
\SetKwInOut{Output}{Output}
\SetKwInOut{Parameters}{Parameters}

\newcommand{\bbR}{\mathbb{R}}

\newcommand{\calB}{\mathcal{B}}
\newcommand{\calC}{\mathcal{C}}
\newcommand{\calD}{\mathcal{D}}

\newcommand{\calF}{\mathcal{F}}

\newcommand{\calH}{\mathcal{H}}

\newcommand{\calK}{\mathcal{K}}

\newcommand{\calM}{\mathcal{M}}

\newcommand{\calO}{\mathcal{O}}

\newcommand{\calR}{\mathcal{R}}
\newcommand{\calS}{\mathcal{S}}

\newcommand{\calU}{\mathcal{U}}
\newcommand{\calV}{\mathcal{V}}

\newcommand{\calX}{\mathcal{X}}

\usepackage{amsthm}

\theoremstyle{definition}
\newtheorem{assumption}{Assumption}%[section]
\newtheorem{theorem}{Theorem}

\newtheorem{proposition}[theorem]{Proposition}
 
\newtheorem{definition}{Definition}%[section]

\newtheorem{example}{Example}

\theoremstyle{remark}
\newtheorem{remark}{Remark}

\DeclareMathOperator*{\argmin}{arg\,min}
\DeclareMathOperator*{\argmax}{arg\,max}

\begin{document}
\title{Predictive Synthesis of Control Barrier Functions and its Application to Time-Varying Constraints}
\author{Adrian Wiltz, \IEEEmembership{Graduate Student Member, IEEE}, and Dimos V. Dimarogonas, \IEEEmembership{Fellow, IEEE}
\thanks{This work was supported by the ERC Consolidator Grant LEAFHOUND, the Swedish Research Council, and the Knut and Alice Wallenberg Foundation. The authors are with the Division of Decision and Control Systems, KTH Royal Institute of Technology, SE-100 44 Stockholm, Sweden {\tt\small \{wiltz,dimos\}@kth.se}.}}
%\thanks{The authors are with the Division of Decision and Control Systems, KTH Royal Institute of Technology, SE-100 44 Stockholm, Sweden {\tt\small \{wiltz,dimos\}@kth.se}.}}

\maketitle

\begin{abstract}
This paper presents a systematic method for synthesizing a Control Barrier Function (CBF) that encodes predictive information into a CBF. Unlike other methods, the synthesized CBF can account for changes and time-variations in the constraints even when constructed for time-invariant constraints. This avoids recomputing the CBF when the constraint specifications change. The method provides an explicit characterization of the extended class~$ \calK_{e} $ function $ \alpha $ that determines the dynamic properties of the CBF, and $ \alpha $ can even be explicitly chosen as a design parameter in the controller synthesis. The resulting CBF further accounts for input constraints, and its values can be determined at any point without having to compute the CBF over the entire domain. The synthesis method is based on a finite horizon optimal control problem inspired by Hamilton-Jacobi reachability analysis and does not rely on a nominal control law. The synthesized CBF is time-invariant if the constraints are. The method poses mild assumptions on the controllability of the dynamic system and assumes the knowledge of at least a subset of some control invariant set. The paper provides a detailed analysis of the properties of the synthesized CBF, including its application to time-varying constraints. A simulation study applies the proposed approach to various dynamic systems in the presence of time-varying constraints. The paper is accompanied by an online available parallelized implementation of the proposed synthesis method.
\end{abstract}


%%%%%%%%%%%%%%%%%%%%%%%%%%%%%%%%%%%%%%%%%%%%%%%%%%%%%%%%%%%%%%%%%%%%%%%%%%%%%%%%
% INTRODUCTION

\section{Introduction}

The dynamic capabilities of a system with respect to a given state constraint can be effectively characterized by a Control Barrier Function (CBF). Thereby, CBFs constitute an important control theoretic tool for ensuring constraint satisfaction and the systematic construction of safety filters~\cite{Wieland2007,Ames2017}.  Having their origin in the optimization literature as barrier functions~\cite{Wills2004,Wright2005}, CBFs became a by now well-established and wide-spread tool within the domain of control~\cite{Hobbs2023,Wabersich2023a}. The application of CBFs has been explored in a wide range of areas such as vehicle control~\cite{Ames2014,Son2019}, vehicle coordination~\cite{Xiao2019a,He2021,Frauenfelder2023}, the control of vessels~\cite{Marley2021a,Peng2023,Haraldsen2024} and underwater vehicles~\cite{Oezkahraman2020}, for air- and spacecrafts~\cite{Barry2012,Breeden2022a,Molnar2025}, to handle sensor limitations~\cite{Huang2022}, as well as in robotics~\cite{ShawCortez2021,Patterson2024,Rauscher2016,Ferraguti2020,FernandezAyala2023}. Moreover, CBFs have been suggested for handling various classes of spatio-temporal constraints~\cite{Lindemann2019,Xiao2021c,Wiltz2022a,Yu2024}. Once a CBF has been found, the design of a controller ensuring constraint satisfaction is rather straightforward. A challenge, however, remains the systematic synthesis of CBFs --- in particular for systems with input constraints or subject to time-varying state constraints. 

%The problem of deriving a CBF is stated as follows: Let a dynamic system
%\begin{align*}
%	\dot{x} = f(x,u)
%\end{align*}
%and a state constraint
%\begin{align*}
%	x\in\calH \coloneq \{x \, | \, h(x)\geq0\}
%\end{align*}
%be given. Then a Lipschitz-continuous function $ b $ shall be derived such that $ \calC\coloneq\{ x \, | \, b(x)\geq 0\}\subseteq\calH $ is a control-invariant subset of $ \calH $ with respect to the dynamic system. In addition, a certain ascend condition on $ b $ yet to be specified must be satisfied. Such a function~$ b $ is then called a CBF.

The problem of deriving a CBF is stated as follows: Let a dynamic system $ \dot{x} = f(x,u) $ and a state constraint $ x\in\calH \coloneq \{x \, | \, h(x)\geq0\} $ be given. Then, a Lipschitz-continuous function $ b $ shall be derived such that $ \calC\coloneq\{ x \, | \, b(x)\geq 0\}\subseteq\calH $ is a control-invariant subset of $ \calH $ with respect to the dynamic system. In addition, a certain ascend condition on $ b $ yet to be specified must be satisfied. Such a function~$ b $ is called a CBF.

If the dynamics are locally controllable on the boundary of $ \calH $ and sufficiently large control inputs $ u $ are admitted, then the design of a CBF is straightforward. In particular, a CBF is then directly given by $ h $. Beyond such favorable cases, CBFs are, like all other value functions as well, notoriously hard to compute. This is especially true for systems with weak controllability properties, such as systems that are \emph{not} locally controllable, or that are subject to input constraints. Then more sophisticated synthesis methods are needed. We start by providing a survey on available methods.

\subsection{Related Work}

The synthesis of CBFs has developed into a thriving research branch in its own right and a wide variety of construction methods have been proposed, each of them with its own advantages. The proposed methods can be roughly subdivided into analytical and numerical approaches, even though some of them combine both analytical and numerical elements. 

At first, we review analytical construction approaches. As previously discussed, the constraint function $ h $ readily constitutes a CBF in some favorable cases. In less favorable cases, backstepping as known from nonlinear control design may still lead to a CBF~\cite{Nguyen2016,Tan2021,Xiao2021b}. Starting with $ h $, which is then called a high-order CBF, an actual CBF can be derived. Yet it is important to note that with the order of the CBF also its sensitivity with respect to model uncertainties increases. A related, though model-free approach is prescribed performance control (PPC)~\cite{Mehdifar2024,Namerikawa2024}. It is applicable to systems that possess relatively strong controllability properties. Other works develop a systematic analytical CBF construction for particular classes of dynamic systems~\cite{Cortez2022a,Tayal2024}, or augment a function that almost everywhere (a.e.) satisfies the CBF-properties with a logic to handle the remaining critical states~\cite{Marley2021}. In order to account for input constraints, analytical approaches commonly require, if at all possible, a meticulous construction of the CBF.

Often, the explicit consideration of input constraints and more generic classes of dynamic systems are possible with numerical approaches. As such, a sum-of-squares (SOS) approach is taken in~\cite{Xu2018a,Clark2021,Wang2022}. Here, a CBF is determined by solving an optimization problem over a polynomial basis. Approaches based on sum-of-squares are limited to polynomial dynamics. Due to the complexity of the optimization problem, there is no guarantee that a CBF is found even if it exists.

Since most control laws based on CBFs are gradient-based, they result in a reactive behavior. Therefore, if a more anticipatory control performance is required from a CBF-based feedback control law, predictive information needs to be encoded into the CBF upon synthesis. Some control approaches circumvent this problem by combining CBFs with model predictive control (MPC)~\cite{Rosolia2022,Charitidou2022,Wabersich2023}, yet, these assume that a CBF is readily provided. In~\cite{Gurriet2018a}, an a-priori known control-invariant set is extended with finite horizon predictions, however without constructing a CBF. The first works to construct a CBF via predictions were~\cite{Squires2018,Breeden2021}. These approaches simulate the system dynamics controlled by some given nominal control law over an infinite time horizon. By additionally employing an a-priori known CBF,~\cite{Chen2021b} reduces the prediction horizon to a finite one. The hereby resulting CBF is called a backup CBF in the literature. A predictive CBF is proposed in~\cite{Breeden2022} using a finite, though not further specified, prediction horizon. The method is based on a sensitivity analysis by varying a nominal control law. 

The need for nominal control laws can be circumvented by choosing an approach via Hamilton-Jacobi reachability analysis as in~\cite{Choi2021}. Despite the time-invariance of the state constraint, a time-dependent barrier function is obtained here. Only with an infinite number of iterations, a time-invariant barrier function can be asymptotically obtained. Its zero super-level set, however, is the largest control-invariant set that ensures the satisfaction of the original state constraint. 

The last category to point out among the numerical CBF synthesis approaches are the learning based ones, see~\cite{Chang2019,Lindemann2024a,Dawson2023} and references therein. Based on the observed trajectories of a dynamical system, %which are rich in information on its dynamical properties, 
a neural network can be trained to approximate a CBF. Learning-based approaches inherently take input constraints into account and the class~$ \calK $ in the ascend condition of the CBF is often obtained along with the learnt CBF. A challenge of such approaches is the generation of the large amount of trajectories needed for the training as well as the verification of the learnt function as a CBF.

A major problem in the CBF synthesis, also within the above mentioned literature, is that CBFs are derived for a particular (static) constraint. Any change in the constraint requires a recomputation of the CBF, which is usually computationally expensive. At the same time, it is important to note that a CBF constructed for a given dynamic system and a given state constraint is far from being unique. 

In this paper, we deliberately take advantage of this freedom in order to construct a CBF with various favorable properties. The here proposed construction method yields a CBF that encodes predictive information and accounts for potential time-variations in the state constraint. Thereby, the resulting CBF does not require an expensive recomputation if the state constraint varies over time but can be adapted. In our follow-up paper~\cite{Wiltz2024b}, we furthermore leverage the here proposed approach to the synthesis of CBFs for equivariant systems~\cite{Field1970,Mahony2020}. Our particular contributions are as follows.

\subsection{Contributions}

Given a dynamic system, a state constraint, and a control-invariant set (or one of its subsets) satisfying the state constraint, our method synthesizes a CBF that encodes predictive information and has the following favorable properties:
\begin{itemize}
	\item The CBF accounts for time-varying constraints. While the synthesized CBF, in the sequel denoted by $ b $, is time-invariant, the function 
	\begin{align*}
		b(x) + \lambda(t)
	\end{align*}
	is guaranteed to be still a CBF for any function $ \lambda $ that varies within certain bounds. These bounds are a design parameter to the proposed synthesis method. Following the terminology in~\cite{Wiltz2024a}, the constructed CBF is \emph{shiftable}. To the best of our knowledge, no other available CBF synthesis method provides this property.
	\item The CBF can be determined on any domain containing its zero super-level set. Most synthesis methods in the literature yield only CBFs on a domain equal to their zero super-level set. 
	\item Our synthesis method yields an explicit characterization of the extended class~$ \calK_{e} $ function in the ascend condition along with the corresponding CBF. It is furthermore indifferent to the relative degree of the constraint.  
	\item Our method allows to compute the numeric values of the CBF at some given state without the need to compute the entire CBF. This is particularly advantageous for exploiting the equivariances of some dynamics in the CBF construction as detailed in our follow-up work~\cite{Wiltz2024b}.
%	\item  Our method allows to compute the CBF values exactly over a grid of points. The overall CBF can be then closely approximated by interpolation or fitting a function to the set of computed values. 
\end{itemize}
Beyond these, our synthesized CBF explicitly accounts for input constraints and can handle a general class of nonlinear dynamics which it has in common with other predictive CBF synthesis methods. However, our synthesis method does not rely on an auxiliary nominal controller in contrast to other predictive methods. In contrast, our method is rather inspired by Hamilton-Jacobi reachability as in~\cite{Choi2021}.

A preliminary version of our synthesis method has been presented in~\cite{Wiltz2023a}. However, the domain of the CBF constructed there was confined to the zero super-level set of the CBF. Moreover, our earlier paper did not account for time-varying constraints. Additionally, we accompany the theoretic results in this paper with detailed implementation remarks, and an elaborate simulation study for various dynamic systems subject to static and time-varying constraints. Furthermore, this paper comes along with a Python toolbox implementing the proposed synthesis method. It supports the parallelized CBF computation on multiple cores. 

\subsection{Outline}

The remainder is structured as follows. Section~\ref{sec:preliminaries} introduces some preliminaries including the definition of a CBF in the Dini sense. Section~\ref{sec:problem setting} states fundamental assumptions on a control invariant set and on the required controllability properties before stating the objective of the paper formally. Section~\ref{sec:pointwise predicitve cbf construction} presents our CBF synthesis method, and Section~\ref{sec:properties} provides a detailed analysis and discussion of its properties. Section~\ref{sec:implementation remarks} remarks on the implementation of our method. Finally, in Section~\ref{sec:simulations}, we apply our method to various dynamic systems and time-varying constraints, and present numerical results. Section~\ref{sec:conclusion} draws a conclusion. 

\subsection{Notation}

Let $ \calX\subseteq\bbR^{n} $, $ x\in\bbR^{n} $. We denote sets by calligraphic upper case letters, while trajectories $ \bm{x}:\bbR \rightarrow \calX $ are denoted by boldface lower case letters. The set of all trajectories $ \bm{x} $ defined on $ [t_{1},t_{2}] $ is denoted by $ \bm{\calX}_{[t_{1},t_{2}]} $, and $ \bm{\calX} $ is used for brevity whenever the interval of definition is clear from the context. The complement and boundary of $ \calX $ are denoted by $ \calX^{c} $ and $ \partial\calX $, respectively, and the Euclidean norm and Hausdorff distance by $ ||\cdot|| $ and $ d_{H}(x,\calX) \coloneq \inf_{x'\in\calX}||x - x'|| $. A ball around $ x_{0} $ with radius $ r $ is defined as $ \calB_{r}(x_{0}) \coloneq \{x \, | \, ||x_{0}-x||< r\} $. For two sets $ \calX_{1},\calX_{2}\subseteq\bbR^{n} $, the Minkowski sum is defined as $ \calX_{1}\oplus\calX_{2} = \{x_{1}+x_{2} \, | \, x_{1}\in\calX_{1}, \, x_{2}\in\calX_{2}\} $, and the Pontryagin difference as $ \calX_{1} \ominus \calX_{2} \coloneq \{x_{1} \in \bbR^{n} \, | \, x_{1} + x_{2} \in\calX_{1}, \; \forall x_{2}\in\calX_{1} \} $. A class~$ \calK $ function is defined as a continuous, strictly increasing function $ \alpha: \bbR_{\geq0} \rightarrow \bbR_{\geq 0} $ with $ \alpha(0) = 0 $; if the function is defined on $ \bbR $ as $ \alpha: \bbR\rightarrow\bbR $, then it is called an extended class~$ \calK_{e} $ function. At last, if a property holds everywhere except on a set of measure zero, we say that it holds almost everywhere~(a.e.).

%%%%%%%%%%%%%%%%%%%%%%%%%%%%%%%%%%%%%%%%%%%%%%%%%%%%%%%%%%%%%%%%%%%%%%%%%%%%%%%%
% PRELIMINARIES

\section{Preliminaries}
\label{sec:preliminaries}

We consider the dynamic system
\begin{align}
	\label{eq:dynamics}
	\dot{x} = f(x,u), \qquad x(0) = x_{0}
\end{align}
where $ x, x_{0} \in\bbR^{n} $, $ u\in\calU\subseteq\bbR^{m} $, and $ f: \bbR^{n}\times\calU \rightarrow \bbR^{n} $ is Lipschitz continuous in both of its arguments. The system is subject to the state constraint
\begin{align}
	\label{eq:state constraint}
	x\in\calH \coloneq \{x \, | \, h(x)\geq0\}
\end{align}
where $ h: \bbR^{n} \rightarrow \bbR $ is a Lipschitz continuous function. For a given input trajectory $ \bm{u}: \bbR_{\geq0} \rightarrow \bm{\calU} $, continuous a.e., the solution to~\eqref{eq:dynamics} up to some time $ T $ is given by $ \bm{\varphi}: [0,T] \rightarrow \bbR^{n} $ where $ \bm{\varphi}(t;x_{0},\bm{u}) \coloneq x_{0} + \int_{0}^{T} f(x(\tau),\bm{u}(\tau)) d\tau $. The forward completeness of $ \bm{\varphi} $ is assumed for the considered input trajectory $ \bm{u} $. We call a set $ \calS $ \emph{forward control invariant} with respect to system~\eqref{eq:dynamics} if there exist $ \bm{u}\in\bm{\calU}_{[0,\infty)} $ such that $ \bm{\varphi}(t;x_{0},\bm{u})\in\calS $ for all $ t\geq 0 $. Furthermore, we call $ \calS $ \emph{forward invariant} under input $ \bm{u}\in\bm{\calU}_{[0,\infty)} $ with respect to~\eqref{eq:dynamics} if $ \bm{\varphi}(t;x_{0},\bm{u})\in\calS $ for all $ t\geq0 $.

\subsection{Control Barrier Functions in the Dini Sense}

Control Barrier Functions (CBF) have been introduced as the system theoretic analogue to Contol Lyapunov Functions (CLF) for forward set invariance~\cite{Wieland2007,Ames2017}. While these works introduce them as differentiable functions, requiring differentiability can be limiting. The need for non-differentiable CBFs arises in the context of constraints with non-differentiable outer bounds (e.g. box constraints), or due to certain dynamics as it is the case for CLFs~\cite{Brockett1983,Clarke2011}. For this reason, we state CBFs more generally in terms of the Dini derivative analogously to CLFs in the Dini sense~\cite{Clarke2011,Sontag1983}.

\begin{definition}[CBF in the Dini Sense]
	\label{def:cbf dini}
	Consider $ \calD\subseteq\bbR^{n} $ and a locally Lipschitz continuous function $ b: \bbR^{n} \rightarrow \bbR $ such that $ \calC $ defined as 
	\begin{align}
%		\label{eq:calC}
		\calC \coloneq \{ x \, | \, b(x)\geq 0 \}
	\end{align}
	is compact and it holds $ \calC\subseteq\calD\subseteq\bbR^{n} $. We call such $ b $ a \emph{CBF in the Dini sense} on $ \calD $ with respect to~\eqref{eq:dynamics} if there exists an extended class~$ \calK_{e} $ function $ \alpha $ such that for all~$ x\in\calD $
	\begin{align}
		\label{eq:def cbf dini}
		\sup_{u\in\calU} \left\{ db(x;f(x,u)) \right\} \geq -\alpha(b(x))
	\end{align}
	where $ d\phi(x;v) $ with $ \phi: \bbR^{n} \rightarrow \bbR $ locally Lipschitz continuous denotes the \emph{Dini derivative} at $ x $ in the direction of $ v $ as
	\begin{align*}
%		\label{eq:dini}
		d\phi(x;v) \coloneq \liminf_{\sigma \downarrow 0}\frac{\phi(x+\sigma v) - \phi(x)}{\sigma}.
	\end{align*}
\end{definition}
\begin{remark}
	In this paper, we explicitly allow for a domain $ \calD $ that is larger than $ \calC $. This is in contrast to our earlier work~\cite{Wiltz2023a}.
\end{remark}

For convenience, we define the other superlevel sets of $ b $ as 
\begin{align*}
	\calC_{\lambda} \coloneq \{ x \, | \, b(x) \geq -\lambda \}
\end{align*}
where $ \lambda\in\bbR $. Next, we briefly show that CBFs in the Dini sense characterize control inputs that render $ \calC $ forward invariant; this is analogous to the differentiable case. The subsequent result is a consequence of the Comparison Lemma.

\begin{theorem}
	Let $ \bm{u}\in\bm{\calU}_{[0,T]} $ be continuous a.e. with the corresponding state trajectory $ \bm{x}(t) \coloneq \bm{\varphi}(t;x_{0},\bm{u}) $ starting in some initial state $ x_{0}\in\calC $. If 
	\begin{align*}
		db(\bm{x}(t);f(\bm{x}(t),\bm{u}(t)))\geq-\alpha(b(\bm{x}(t))) \qquad \forall t\in[0,T],
	\end{align*}
	then $ \calC $ is forward invariant such that $ \bm{x}(t)\in\calC $ for all $ t\in[0,T] $.
\end{theorem}

\begin{proof}
	Let the intervals, where $ \bm{u} $ is continuous, be w.l.o.g. given as $ [\tau_{i},\tau_{i+1}) $ where $ \tau_{i}\in\{\tau_{i}\}_{i=1,\dots,N\!-\!1} $ with 
	\begin{align*}
		\tau_{0} = 0 < \dots < \tau_{i} < \tau_{i+1} < \dots < \tau_{N} = T.
	\end{align*} 
	Next, assuming that $ b(\bm{x}(\tau_{i}))\geq0 $, it follows together with $ db(\bm{x}(t);f(\bm{x}(t),\bm{u}(t)))\geq-\alpha(b(\bm{x}(t))) $ from the Comparison Lemma~\cite[Lemma~3.4]{Khalil2002} that $ b(\bm{x}(t)) \geq 0 $ for all $ t\in[\tau_{i},\tau_{i+1}) $. Due to the continuity of $ b $ and $ \bm{x} $, we further obtain 
	\begin{align*}
		b(\bm{x}(\tau_{i+1})) = \liminf_{\tau\uparrow\tau_{i+1}} b(\bm{\varphi}(\tau;\bm{x}(\tau_{i}),\bm{u})) \geq 0.
	\end{align*}
	Since $ x_{0}\in\calC $ and thus $ b(x_{0}) = b(\bm{x}(\tau_{0})) \geq 0 $, it follows inductively that $ b(\bm{x}(t)) \geq 0 $ and therefore $ \bm{x}(t)\in\calC $ for all $ t\in[0,T] $. 
	% Do we need a note here on why the partitioning of the time horizon is w.l.o.g.???
\end{proof}

\subsection{Reachability and Controllability}
A state $ x_{1} $ is called \emph{T-reachable} from $ x_{0} $ under dynamics~\eqref{eq:dynamics} if $ \bm{\varphi}(T;x_{0},\bm{u}) = x_{1} $ for some bounded measurable input trajectory $ \bm{u}\in\bm{\calU}_{[0,T]} $. We define the set of all such points as $ \calR_{T}(x_{0}) \coloneq \{x_{1} \, | \, \exists \bm{u}\in\bm{\calU}_{[0,T]}: \bm{\varphi}(T; x_{0},\bm{u}) = x_{1} \} $. System~\eqref{eq:dynamics} is \emph{controllable }on $ \calM\subseteq\bbR^{n} $ if $ \calM \subseteq \bigcup_{t\in[0,\infty)} \calR_{t}(x_{0}) $ for all $ x_{0}\in\calM $~\cite{Hermann1977}. Moreover, we call system~\eqref{eq:dynamics} \emph{locally-locally controllable} on $ \calM\subseteq\bbR^{n} $~\cite{Haynes1970} if there exist $ \varepsilon \geq \delta_{\varepsilon} > 0 $ such that for any state $ x_{f}\in \calB_{\delta_{\varepsilon}}(x_{0}) $ there exists a $ \bm{u}\in\bm{\calU} $ and $ t\geq0 $ such that $ \bm{\varphi}(t;x_{0},\bm{u}) = x_{f} $ and $ \bm{\varphi}(\tau;x_{0},\bm{u}) \in \calB_{\varepsilon}(x_{0})$ for all $ \tau\in[0,t] $ (``any state in a $ \delta_{\varepsilon} $-neighborhood can be reached without leaving a certain $ \varepsilon $-neighborhood'').

%%%%%%%%%%%%%%%%%%%%%%%%%%%%%%%%%%%%%%%%%%%%%%%%%%%%%%%%%%%%%%%%%%%%%%%%%%%%%%%%
% MAIN RESULTS

\section{Problem Setting}
\label{sec:problem setting}

Let us consider the constraint set 
\begin{align*}
	\calH \coloneq \{x \, | \, h(x)\geq0\}
\end{align*}
where $ h $ is a Lipschitz continuous function. The CBF synthesis method presented in this paper is based on the idea that if a system is initialized with $ x_{0} $ sufficiently far from the boundary of constraint set~$ \calH $, then it can be expected that there exists an input trajectory which ensures that the system state stays within $ \calH $ for all times. This property holds, for example, for all locally-locally controllable systems, but also for any more general system that possesses a forward control invariant set contained in~$ \calH $. The particular premises, under which a CBF is constructed in the sequel, are formalized by the following assumptions. These parallel the setting in our previous work~\cite{Wiltz2023a}.

\subsection{Assumption on Forward Control Invariant Sets}
\label{subsec:problem setting assumptions control invarinat set}
First, we assume the existence of a forward control invariant set $ \calV $ which does not need to be explicitly known. Instead, we assume that only one of its subsets, denoted by $ \calF $, is known. 

\begin{assumption}
	\label{ass:setF}
	There exists a forward control invariant subset $ \calV\subset\calH $ where $ \calH \coloneq \{x\in\bbR^{n}\, | \, h(x) \geq 0 \} $ such that $ h(x)\geq \delta $ for all $ x\in\calV $ and some $ \delta > 0 $. While $ \calV $ is not required to be explicitly known, we assume that a subset $ \calF\subseteq\calV $ is known.
\end{assumption}

\begin{figure}[t]
	\centering
	\def\svgwidth{0.67\columnwidth}
	%% Creator: Inkscape 1.3.2 (091e20e, 2023-11-25, custom), www.inkscape.org
%% Accompanies image file 'setF_new.pdf' (pdf, eps, ps)
\begingroup%
  \makeatletter%
  \providecommand\color[2][]{%
    \errmessage{(Inkscape) Color is used for the text in Inkscape, but the package 'color.sty' is not loaded}%
    \renewcommand\color[2][]{}%
  }%
  \providecommand\transparent[1]{%
    \errmessage{(Inkscape) Transparency is used (non-zero) for the text in Inkscape, but the package 'transparent.sty' is not loaded}%
    \renewcommand\transparent[1]{}%
  }%
  \providecommand\rotatebox[2]{#2}%
  \newcommand*\fsize{\dimexpr\f@size pt\relax}%
  \newcommand*\lineheight[1]{\fontsize{\fsize}{#1\fsize}\selectfont}%
  \ifx\svgwidth\undefined%
    \setlength{\unitlength}{413.59952671bp}%
    \ifx\svgscale\undefined%
      \relax%
    \else%
      \setlength{\unitlength}{\unitlength * \real{\svgscale}}%
    \fi%
  \else%
    \setlength{\unitlength}{\svgwidth}%
  \fi%
  \global\let\svgwidth\undefined%
  \global\let\svgscale\undefined%
  \makeatother%
  \begin{picture}(1,0.40710382)%
    \lineheight{1}%
    \setlength\tabcolsep{0pt}%
    \put(0,0){\includegraphics[width=\unitlength,page=1]{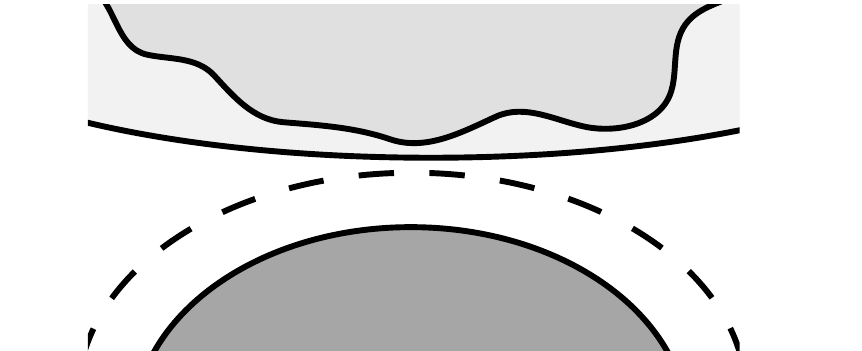}}%
    \put(0.38903693,0.04392397){\color[rgb]{0,0,0}\makebox(0,0)[lt]{\lineheight{1.25}\smash{\begin{tabular}[t]{l}{\small$h(x)<0$}\end{tabular}}}}%
    \put(0.38474308,0.15796462){\color[rgb]{0,0,0}\makebox(0,0)[lt]{\lineheight{1.25}\smash{\begin{tabular}[t]{l}{\small$h(x)>0$}\end{tabular}}}}%
    \put(0,0){\includegraphics[width=\unitlength,page=2]{setF_new.pdf}}%
    \put(0.05876518,0.17413546){\color[rgb]{0,0,0}\makebox(0,0)[lt]{\lineheight{1.25}\smash{\begin{tabular}[t]{l}$h(x)=\delta$\end{tabular}}}}%
    \put(0.46402124,0.33398824){\color[rgb]{0,0,0}\makebox(0,0)[lt]{\lineheight{1.25}\smash{\begin{tabular}[t]{l}$\mathcal{F}$\end{tabular}}}}%
    \put(0.15742607,0.2813495){\color[rgb]{0,0,0}\makebox(0,0)[lt]{\lineheight{1.25}\smash{\begin{tabular}[t]{l}$\mathcal{V}$\end{tabular}}}}%
  \end{picture}%
\endgroup%

	\caption{Illustration of Assumption~\ref{ass:setF}.}
	\label{fig:setF}
	\vspace{-\baselineskip}
\end{figure}

The assumption is illustrated in Figure~\ref{fig:setF}. Often, the construction of $ \calF $ is more straightforward than that of $ \calV $, and an intuitive understanding of the system dynamics can be taken as a starting point. 

\begin{figure}[t]
	\centering
	\begin{subfigure}[b]{0.4\columnwidth}
		\centering
		\def\svgwidth{1\columnwidth}
		%% Creator: Inkscape 1.1.1 (3bf5ae0d25, 2021-09-20), www.inkscape.org
%% Accompanies image file 'bicycle.pdf' (pdf, eps, ps)
\begingroup%
  \makeatletter%
  \providecommand\color[2][]{%
    \errmessage{(Inkscape) Color is used for the text in Inkscape, but the package 'color.sty' is not loaded}%
    \renewcommand\color[2][]{}%
  }%
  \providecommand\transparent[1]{%
    \errmessage{(Inkscape) Transparency is used (non-zero) for the text in Inkscape, but the package 'transparent.sty' is not loaded}%
    \renewcommand\transparent[1]{}%
  }%
  \providecommand\rotatebox[2]{#2}%
  \newcommand*\fsize{\dimexpr\f@size pt\relax}%
  \newcommand*\lineheight[1]{\fontsize{\fsize}{#1\fsize}\selectfont}%
  \ifx\svgwidth\undefined%
    \setlength{\unitlength}{256.26495556bp}%
    \ifx\svgscale\undefined%
      \relax%
    \else%
      \setlength{\unitlength}{\unitlength * \real{\svgscale}}%
    \fi%
  \else%
    \setlength{\unitlength}{\svgwidth}%
  \fi%
  \global\let\svgwidth\undefined%
  \global\let\svgscale\undefined%
  \makeatother%
  \begin{picture}(1,0.81626357)%
    \lineheight{1}%
    \setlength\tabcolsep{0pt}%
    \put(0,0){\includegraphics[width=\unitlength,page=1]{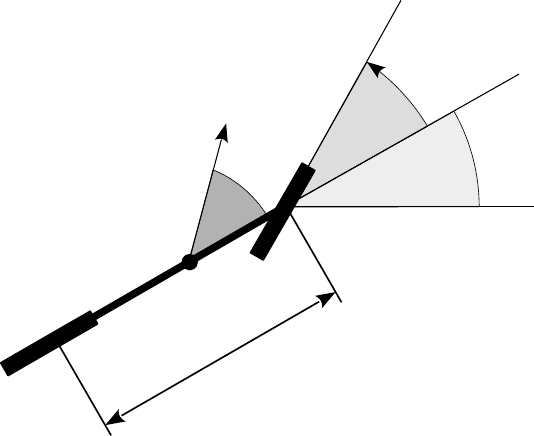}}%
    \put(0.39607649,0.40454786){\color[rgb]{0,0,0}\makebox(0,0)[lt]{\lineheight{1.25}\smash{\begin{tabular}[t]{l}$\beta$\end{tabular}}}}%
    \put(0.64354382,0.56033633){\color[rgb]{0,0,0}\makebox(0,0)[lt]{\lineheight{1.25}\smash{\begin{tabular}[t]{l}$\zeta$\end{tabular}}}}%
    \put(0.78059458,0.4530023){\color[rgb]{0,0,0}\makebox(0,0)[lt]{\lineheight{1.25}\smash{\begin{tabular}[t]{l}$\psi$\end{tabular}}}}%
    \put(0.43136955,0.06876114){\color[rgb]{0,0,0}\makebox(0,0)[lt]{\lineheight{1.25}\smash{\begin{tabular}[t]{l}$L$\end{tabular}}}}%
    \put(0.34575181,0.234386){\color[rgb]{0,0,0}\makebox(0,0)[lt]{\lineheight{1.25}\smash{\begin{tabular}[t]{l}$C$\end{tabular}}}}%
    \put(0.32587123,0.43125159){\color[rgb]{0,0,0}\makebox(0,0)[lt]{\lineheight{1.25}\smash{\begin{tabular}[t]{l}$v$\end{tabular}}}}%
    \put(0,0){\includegraphics[width=\unitlength,page=2]{bicycle.pdf}}%
  \end{picture}%
\endgroup%

		\caption{}
		\label{fig:kinematic bicycle model}
	\end{subfigure}
	\hfill
	\begin{subfigure}[b]{0.52\columnwidth}
		\centering
		\def\svgwidth{1\columnwidth}
		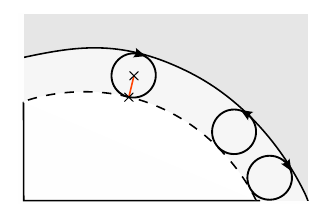
		\caption{}
		\label{fig:consrtuction safe set bicycle}
	\end{subfigure}
	\caption{Kinematic bicycle model: (a) schematic sketch; (b) construction of a set with its viable points via turning radius~$ r $.}
	\vspace{-\baselineskip}
\end{figure}

\begin{example}
	\label{exmp:bicycle1}
	The kinematic model of a vehicle modeled as a bicycle \cite{Wang2001} (see Fig.~\ref{fig:kinematic bicycle model}) is given as %$ \dot{x} = v \cos(\psi + \beta(\zeta)) $, $ \dot{y} = v \sin(\psi + \beta(\zeta)) $, $ \dot{\psi} = \frac{v \cos(\beta(\zeta))\tan(\zeta)}{L} $
	\begin{subequations}
		\label{eq:bicycle model}
		\begin{align}
				\label{seq:bicycle xdot}
				\dot{x} &= v \cos(\psi + \beta(\zeta)) \\
				\label{seq:bicycle ydot}
				\dot{y} &= v \sin(\psi + \beta(\zeta)) \\
				\label{seq:bicycle psidot}
				\dot{\psi} &= \frac{v \cos(\beta(\zeta))\tan(\zeta)}{L} 
			\end{align}
	\end{subequations}
	where $ \beta(\zeta) = \arctan(\frac{1}{2}\tan(\zeta)) $. The position of the center of mass $ C $ is denoted by the states $ \mathbf{x}_{\text{pos}}=[x, y]^{T} $, and the vehicle's orientation by $ \psi $; inputs are velocity~$ v $ and steering angle~$ \zeta $. The vector of the system states is denoted by $ \mathbf{x} = [x,y,\psi]^{T} $. The vehicle is also subject to input constraints $ 0 < v_{\text{min}}\leq v\leq v_{\text{max}} $ and $ |\zeta|\leq \zeta_{\text{max}} $. The minimum turning radius can be directly obtained from the dynamics as $ R=\frac{L}{\cos(\beta(\zeta_{\text{max}}))\tan(\zeta_{\text{max}})} $.
	%	\begin{align*}
		%		r=\frac{L}{\cos(\beta(\zeta_{\text{max}}))\tan(\zeta_{\text{max}})}.
		%	\end{align*}
	We let the vehicle move in a plane with an obstacle as shown in Figure~\ref{fig:consrtuction safe set bicycle}. The obstacle is specified as a set $ \calH^{c}\subset\bbR^{n} $. Correspondingly, we define $ h(\mathbf{x}) = d_{H}(\mathbf{x},\calH^{c}) $.
	By geometrical considerations, as shown in Fig.~\ref{fig:consrtuction safe set bicycle}, the set $ \calF = \lbrace \mathbf{x} \, | \, h(\mathbf{x}) \geq \delta+2R \rbrace $ is determined as a subset of the control-invariant set $ \calV $. It is however important to note that $ \calF $ is not forward control invariant. While for the construction  of $ \calF $ it is sufficient to consider the vehicle's position, the construction of $ \calV $ also requires orientation $ \psi $.
\end{example} 

In the example, we exploit the fact that a bicycle can always return to its initial state by moving on a circle. More generally, this relates to the following observation. 
\begin{proposition}
	\label{prop:setF}
	A set $ \calF $ is a subset of a forward control invariant set if for any state $ x_{0}\in\partial\calF $ there exists a control input $ \bm{u}\in\bm{\calU} $ such that $ \bm{\varphi}(t_{f};x_{0},\bm{u})\in\calF $ for some $ t_{f}\geq 0 $.
\end{proposition}
\begin{proof}
	For a given $ x_{0}\in\partial\calF $, let there be a control input $ \bm{u}_{x_{0}}\in\bm{\calU} $ and a final time $ t_{f,x_{0}} \geq 0 $ such that $ \bm{\varphi}(t_{f,x_{0}};x_{0},\bm{u}_{x_{0}})\in\calF $; their existence is a consequence of the premises of the proposition. Then, note that 
	\begin{align*}
		\calV \coloneq \bigcup_{x_{0}\in\partial\calF} \{ x \, | \, x = \bm{\varphi}(t;x_{0},\bm{u}_{x_{0}}), \; t\in[0,t_{f,x_{0}}] \} \cup \calF
	\end{align*}
	is forward control invariant and $ \calF\subseteq \calV $.
\end{proof}
This however is not a necessary condition, as any trajectory through $ x_{0} $ might converge to some point or limit cycle outside of $ \calF $. For the construction of a set $ \calF $ satisfying Assumption~\ref{ass:setF}, it is of particular interest if any such trajectory starting at an arbitrary $ x_{0}\in\partial\calF $ can be bounded into an $ \varepsilon $-neighborhood of~$ x_{0} $. This gives rise to the following sufficient conditions. % for $ \calF $ that satisfy Assumption~\ref{ass:setF}.

\begin{proposition}
	\label{prop:setF eps}
	The set $ \calF \coloneq \{x \,|\, h(x)\geq\delta\}\ominus B_{\varepsilon}(0) $ for some $ \varepsilon>0 $ is a subset of a forward control invariant set~$ \calV $ that satisfies Assumption~\ref{ass:setF} if at least one of the subsequent conditions hold:
	\begin{enumerate}[label=(\arabic*)]
		\item System~\eqref{eq:dynamics} is locally-locally controllable on $ \partial\calF $ with constants $ \varepsilon \geq \delta_{\varepsilon} > 0 $. \label{cond:prop:set F eps locally-locally controllable}
		\item For any state $ x_{0}\in\partial\calF $, there exists a control input $ \bm{u}\in\bm{\calU} $ such that $ \bm{\varphi}(t;x_{0},\bm{u})\in B_{\varepsilon}(x_{0}) $ for all $ t\geq 0 $ (e.g. $ B_{\varepsilon}(x_{0}) $ contains an attractor, or $ \bm{\varphi} $ converges to a periodic solution in $ B_{\varepsilon}(x_{0}) $). \label{cond:prop:set F eps ball around x0}
		\item For any state $ x_{0}\in\partial\calF $, there exists a control input $ \bm{u}\in\bm{\calU} $ such that $ \bm{\varphi}(t_{f};x_{0},\bm{u})\in\calF $ for some $ t_{f}\geq 0 $ and $ \bm{\varphi}(\tau;x_{0},\bm{u})\in\calF\oplus B_{\varepsilon}(0) $ for all $ \tau\in[0,t_{f}] $. \label{cond:prop:set F eps nbhd of Fhat}
	\end{enumerate}
	Also, any subset of $ \calF $ satisfies Assumption~\ref{ass:setF}.
\end{proposition}
\begin{proof}
	The proof can be found in the appendix.
\end{proof}

This proposition is of practical relevance. For instance, the construction of $ \calF $ in Example~\ref{exmp:bicycle1} is based on condition~\ref{cond:prop:set F eps ball around x0} of the proposition. More generally, the proposition points out that the knowledge on a system's locally-locally controllability or on the existence of attractors and periodic solutions are valuable properties that can be exploited for the construction of $ \calF $.

\subsection{Assumption on Controllability and Reachability}
\label{subsec:problem setting assumptions controllability}

We call a trajectory starting at some point $ x_{0} $ \emph{viable} if there exists an input trajectory $ \bm{u}:\bbR_{\geq 0} \rightarrow \calU $ such that $ \bm{\varphi}(t;x_{0},\bm{u})\in\calH $ for all $ t\geq0 $. By the next assumption, we ensure that the viability of a trajectory can be determined by predicting the trajectories of a system over a finite time horizon.
To this end, we note that a trajectory that ends in $ x_{1}\in\calF $ can be feasibly continued for all times as $ \calF $ is a subset of a forward control invariant set $ \calV\subset\calH $. Thus, the time horizon required in order to decide if a trajectory starting at $ x_{0} $ is viable, is the minimal time $ \tau(x_{0}) $ that it takes to reach $ \calF $. It is formally defined as
\begin{subequations}
	\label{eq:tau_x0}
	\begin{align}
		\label{seq:tau_x0 objective}
		\tau(x_{0}) &\coloneq \min_{\tau\geq 0} \tau \\
		\label{seq:tau_x0 dynamics}
		\text{s.t.} \;\; & \dot{\bm{x}}(t) = f(\bm{x}(t),\bm{u}(t)) \quad \text{(a.e.)}, \\
		\label{seq:tau constraints}
		& \bm{x}(0) = x_{0}, \quad \bm{u}(t)\in\calU, \quad \bm{x}(\tau)\in\calF. 
	\end{align}
\end{subequations}
Note that trajectory $ \bm{x} $ is not required to stay in $ \calH $ for all times. Here, the states $ x_{0}\in\calD $ are of particular interest as a CBF on the domain $ \calD $ shall be constructed. While for $ x_{0}\in\calF $ it clearly holds $ \tau(x_{0}) = 0 $, further assumptions are required to ensure that $ \tau(x_{0}) $ is finite for any other $ x_{0}\in\calD\setminus\calF $. 
To this end, we impose the following controllability assumption.

\begin{assumption}
	\label{ass:controllability}
	Let either of the following statements hold:
	\begin{enumerate}[leftmargin=0.9cm]
		\item[A\ref{ass:controllability}.1] System \eqref{eq:dynamics} is controllable on the closure of $\calF^{\text{c}}$; or
%		 where $\calF^{\text{c}}$ denotes the complement of $ \calF $; or
		\item[A\ref{ass:controllability}.2] For all $ x_{0}\in\calD\setminus\calF $, there exist $ t\geq 0 $ such that $ \calR_{t}(x_{0})\cap\calF \neq \emptyset $. 
	\end{enumerate}
\end{assumption}

\begin{proposition}[\cite{Wiltz2023a}, Proposition~2]
	\label{prop:finite time horizon}
	Let Assumption~\ref{ass:controllability} hold. Then there exists for all $ x_{0}\in\calD\setminus\calF $ a finite time $ \tau(x_{0})\in\bbR_{>0} $ that minimizes~\eqref{eq:tau_x0}.
\end{proposition}

Let us denote the upper bound of $ \tau(\cdot) $ on $ \calD\setminus\calF $ by 
\begin{align}
	\label{eq:time tau def}
	\tau\coloneq\sup_{x\in\calD\setminus\calF}\tau(x)
\end{align}
Upper bounds to $ \tau(\cdot) $ can often be analytically found. An illustrative example on the bicycle model can be found in~\cite{Wiltz2023a}. 

\subsection{Objective}
\label{subsec:problem seeting objective}

Let Assumptions~\ref{ass:setF} and~\ref{ass:controllability} hold, and let a constraint set $ \calH $, a domain $ \calD \supseteq \calH $, and a set $ \calF $ as defined in Assumption~\ref{ass:setF} be given. Then, construct a CBF with respect to~\eqref{eq:dynamics} on domain~$ \calD $ such that its zero-superlevel set $ \calC $ is a subset of constraint set~$ \calH $, i.e., $ \calC \subseteq \calH $. %In this context, note that $ \calD \supseteq \calC $. Thus, $ \calD $ may be a \emph{strict} superset of $ \calC $, which is in contrast to most existing works on CBF synthesis that usually assume $ \calD = \calC $. 
\section{Predictive CBF Synthesis}
\label{sec:pointwise predicitve cbf construction}

We now present our synthesis approach and prove that the synthesized function constitutes a CBF in the Dini sense. Thereafter, we refine the established result in order to relax imposed conditions and to explicitly incorporate the extended class~$ \calK_{e} $ function $ \alpha $ as a design parameter into the synthesis.

\subsection{Synthesis Approach}
\label{subsec:construction}

The CBF is determined pointwise by solving an optimal control problem over a finite prediction horizon $ T $. Thus, the CBF can be computed on parts of its domain without having to compute it on the entire domain. This is, for instance, advantageous when the overall CBF can be induced from its values computed on a subset of its domain as it is the case for equivariant systems~\cite{Wiltz2024b}. 

In order to become more specific, let us choose the prediction horizon $ T $ as $ T \geq \tau $, where $ \tau $ is defined in~\eqref{eq:time tau def}. Moreover, let us define a function~$ H_{T}:\calD \rightarrow \bbR $ as
\begin{subequations}
	\label{eq:finite horizon construction H}
	\begin{align}
		\label{seq:finite horizon construction H max min}
		H_{T}(x_{0}) &\coloneq \max_{\bm{u}(\cdot)\in\bm{\calU}_{[0,T]}} \min_{t\in[0,T]} h(\bm{x}(t)) - \gamma t \\
		\label{seq:finite horizon construction H initial condition}
		\text{s.t.}\;\; &\bm{x}(0)=x_{0},\\
		\label{seq:finite horizon construction H dynamics}
		&\dot{\bm{x}}(s) = f(\bm{x}(s),\bm{u}(s)) \quad (a.e.), \\
		\label{seq:finite horizon construction H input constraint}
		&\bm{u}(s)\in\calU, \qquad \forall s\in[0,T]\\
		\label{seq:finite horizon construction H terminal constraint}
		& \bm{x}(\vartheta)\in\calF, \qquad \text{for some } \vartheta\in[0,T],
	\end{align}
\end{subequations}
where $ \gamma>0 $ is some positive constant. In the sequel, $ H_{T} $ turns out to be a CBF in the Dini sense. For later reference, we denote the input trajectory~$ \bm{u} $ and the times~$ t $ and~$ \vartheta $ that solve optimization problem~\eqref{eq:finite horizon construction H} for initial value $ x_{0} $ by~$ \bm{u}^{\ast}_{x_{0}} $, $ t^{\ast}_{x_{0}} $ and~$ \vartheta^{\ast}_{x_{0}} $. 

The intuition behind optimal control problem~\eqref{eq:finite horizon construction H} is as follows. Let us assume for a moment that $ \gamma = 0 $. In the optimal control problem, we consider a state trajectory $ \bm{x}(\cdot) = \bm{\varphi}(\cdot;x_{0},\bm{u}) $ that starts in $ x_{0} $ and evolves according to some input trajectory $ \bm{u} $ over a time horizon~$ T $ such that~\eqref{seq:finite horizon construction H terminal constraint} is satisfied at some time $ \vartheta\in[0,T] $. The minimization determines that point of time $ t^{\ast}_{x_{0}} $ when the trajectory $ \bm{x} $ takes the smallest value on $ h $, whereas the maximization aims at increasing this value as much as possible with a suitable input trajectory. Thereby for $ \gamma = 0 $, the optimal control problem~\eqref{eq:finite horizon construction H} defines the function $ H_{T} $ such that it can be interpreted as a measure for how close the system state gets to the boundary of set~$ \calH $, or to which extent the state trajectory leaves set $ \calH $ when it evolves over time. Constraint~\eqref{seq:finite horizon construction H terminal constraint} ensures that $ \bm{\varphi}(\cdot; x_{0}, \bm{u}_{x_{0}}^{\ast}) $ can be always feasibly continued within~$ \calH $ even beyond time horizon~$ T $. For $ \gamma = 0 $, \eqref{eq:finite horizon construction H} is identical to the optimal control problem proposed in~\cite{Wiltz2023a}. Next, let us consider $ \gamma $ to be strictly positive, which renders $ -\gamma t $ non-positive on the time-interval~$ [0,T] $ and penalizes the magnitude of $ t^{\ast}_{x_{0}} $. Thereby, it is ensured that the value of $ H_{T} $ increases along trajectory $ \bm{\varphi}(\cdot;x_{0},\bm{u}^{\ast}_{x_{0}}) $. In particular, we formally derive later on in the proof of Theorem~\ref{thm:predictive CBF} that for any $ x_{0}\in\calD\setminus\calF $ it holds
\begin{align*}
	dH_{T}(x;f(x,\bm{u}_{x_{0}}^{\ast})) \bigg|_{x = \bm{\varphi}(t;x_{0},\bm{u}_{x_{0}}^{\ast})} > 0.
\end{align*} 
The latter inequality eventually allows us to establish that $ H_{T} $ constitutes a CBF even on domains $ \calD \supset \calC $. This is in contrast to most existing works on CBF synthesis, which only allow for $ \calD = \calC $. 

Before formally establishing this result, we show that $ H_{T} $ is well-defined and the zero super-level set of $ H_{T} $, in the remainder of the paper denoted by $ \calC \coloneq \{x \, | \, H_{T}(x)\geq 0 \} $, is a subset of the state constraint set $ \calH $, i.e., $ \calC\subseteq\calH $. Thereby, any point in $ \calC $ satisfies state constraint~\eqref{eq:state constraint}. 

\begin{proposition}
	\label{prop:welldefined H_T and C subseteq H}
	Let Assumptions~\ref{ass:setF} and~\ref{ass:controllability} hold. Then for all $ x_{0}\in\calD $, $ H_{T} $ defined in~\eqref{eq:finite horizon construction H} is well-defined, i.e., there exists a solution to~\eqref{eq:finite horizon construction H}. Moreover, $ \calC \subseteq \calH $. 
\end{proposition}
\begin{proof}
	In order to show the first part, consider a state $ x_{0}\in\calD $. As Assumptions~\ref{ass:setF}--\ref{ass:controllability} hold, it follows from Proposition~\ref{prop:finite time horizon} that there exists a finite time horizon $ T $ with $ T\geq\tau $. Thus, by the definition of $ \tau $, there also exists an input trajectory $ \bm{u}^{\ast}_{x_{0}}\in\bm{\calU}_{[0,T]} $ and times $ t^{\ast}_{x_{0}} $ and $ \vartheta^{\ast}_{x_{0}} $ which solve~\eqref{eq:finite horizon construction H}. Thus, $ H_{T} $ is well-defined. 
	
	For the second part, we note that 
	\begin{align*}
		H_{T}(x_{0}) &\stackrel{\eqref{seq:finite horizon construction H max min}}{=} \max_{\bm{u}(\cdot)} \min_{t\in[0,T]} h(\bm{\varphi}(t;x_{0},\bm{u})) - \gamma t \\
		&\leq \max_{\bm{u}(\cdot)} \min_{t\in[0,T]} h(\bm{\varphi}(t;x_{0},\bm{u})) \\
		&\leq \max_{\bm{u}(\cdot)}  h(\bm{\varphi}(0;x_{0},\bm{u}))= h(x_{0}).
	\end{align*}
	Thus, if $ H_{T}(x_{0}) \geq 0 $, then $ h(x_{0})\geq 0 $ and $ \calC\subseteq \calH $.
\end{proof}

\begin{figure}[t]
	\centering
	\def\svgwidth{0.8\columnwidth}
	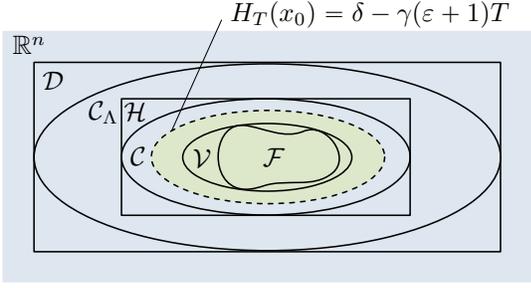
	\caption{Relation of the defined sets. The particular shapes of the sets are only schematic and have no further meaning.
	}
	\label{fig:predictive CBF set relation}
	\vspace{-\baselineskip}
\end{figure}

Thereby, the relation of the so far defined sets can be summarized as
\begin{align*}
	\calF \subseteq \calV \subseteq \calC \subseteq \calH \subseteq \calD \subseteq \bbR^{n}
\end{align*}
which is illustrated in Figure~\ref{fig:predictive CBF set relation}. Now we are ready to prove that $ H_{T} $ constitutes a CBF in the Dini sense.

\begin{theorem}
	\label{thm:predictive CBF}
	Let Assumptions~\ref{ass:setF} and~\ref{ass:controllability} hold, let $ h $ be Lipschitz-continuous and let $ T\geq\tau $. Moreover, let $ H_{T}: \calD\rightarrow\bbR $ be defined by~\eqref{eq:finite horizon construction H} on some domain $ \calD\subseteq\bbR^{n} $ with $ \calH\subset\calD $; the parameter $ \gamma $ is chosen such that
	\begin{align}
		\label{eq:thm:predictive CBF 0}
		\gamma < \frac{\delta}{T}
	\end{align}
	where $ \delta $ is determined as part of Assumption~\ref{ass:setF}. Furthermore, let $ f $ be bounded on $ \calC $ in the sense that for all $ x\in\calC $ there exists a $ u\in\calU $ such that $ ||f(x,u)||\leq M $ for some constant $ M>0 $. If $ H_{T} $ is locally Lipschitz-continuous, then $ H_{T} $ constitutes a CBF in the Dini sense on the domain $ \calD $ with respect to the dynamics~\eqref{eq:dynamics}. 
\end{theorem}

\begin{proof}
	To start with, note that at any $ x_{0}\in\calD $, $ H_{T}(x_{0}) $ is well-defined as by Proposition~\ref{prop:welldefined H_T and C subseteq H}. The value of $ H_{T} $ at $ x_{0} $ is then given by $ H_{T}(x_{0}) = h(\bm{\varphi}(t^{\ast}_{x_{0}};x_{0},\bm{u}_{x_{0}}^{\ast})) $ where $ \bm{\varphi}(\cdot;x_{0},\bm{u}^{\ast}_{x_{0}}):[0,T]\rightarrow \bbR^{n} $ is the state trajectory that starts in $ x_{0} $ and is induced by $ \bm{u}^{\ast}_{x_{0}} $.
	
	In order to prove that $ H_{T} $ is a CBF in the Dini sense, it needs to be shown that for some extended class~$ \calK_{e} $ function $ \alpha $ it holds  
	\begin{align}
		\label{eq:thm:predictive CBF aux 0}
		\sup_{u\in\calU}\left\{dH_{T}(x_{0};f(x_{0},u)) \right\} \geq -\alpha(H_{T}(x_{0}))
	\end{align}
	for all $ x_{0}\in\calD $. To this end, we introduce $ \varepsilon>0 $ such that 
	\begin{align}
		\label{eq:thm:predictive CBF aux 0.1}
		\gamma < \frac{\delta}{(\varepsilon+1)T} < \frac{\delta}{T}
	\end{align}
	which exists due to~\eqref{eq:thm:predictive CBF 0}. As such, we can choose any $ \varepsilon\in(0,\frac{\delta}{\gamma T} - 1) $. From here on, we proceed in two steps: at first, we consider those $ x_{0} $ with $ H_{T}(x_{0}) \leq \delta -\gamma (\varepsilon+1)T $ (marked blue in Figure~\ref{fig:predictive CBF set relation}); thereafter, we consider the remaining $ x_{0} $, namely those with $ H_{T}(x_{0}) > \delta -\gamma (\varepsilon+1)T $ (marked in green). 
	
	\emph{Step 1:} Let $ x_{0} $ be such that $ H_{T}(x_{0})\leq \delta-\gamma (\varepsilon+1)T $. Note that the right-hand side $ \delta-\gamma (\varepsilon+1)T $ is strictly positive due to~\eqref{eq:thm:predictive CBF aux 0.1}. In this way, Step~1 considers all states $ x_{0} $ outside of $ \calC $ ($  x_{0}\notin\calC $), together with those contained in $ \calC $ within in some neighborhood of its boundary (see Figure~\ref{fig:predictive CBF set relation}). 
	Now let us extend input trajectory $ \bm{u}^{\ast}_{x_{0}} $ by an input trajectory $ \bm{u}_{e,x_{0}}\in\bm{\calU}_{[T,\infty)} $ that renders $ \calV $ forward invariant for all times $ t\geq T $. In particular,  
	\begin{align}
		\label{eq:thm:predictive CBF aux 1}
		\bm{u}_{e,x_{0}}^{\ast}(t) \coloneq 
		\begin{cases}
			\bm{u}^{\ast}_{x_{0}}(t) &\text{if } t\in[0,\vartheta^{\ast}_{x_{0}}] \\
			\bm{u}_{e,x_{0}}(t) &\text{if } t\in(\vartheta^{\ast}_{x_{0}},T(\varepsilon+1)]
		\end{cases}
	\end{align}
	where $ \bm{u}_{e,x_{0}}\in\bm{\calU}_{(\vartheta^{\ast}_{x_{0}},T(\varepsilon+1)]} $ such that $ \bm{\varphi}(t;x_{0},\bm{u}_{e,x_{0}}^{\ast})\in\calV $ for all $ t>\vartheta^{\ast}_{x_{0}} $. Since $ \calV $ is forward control invariant, such $ \bm{u}_{e,x_{0}} $ exists. Moreover, we note that
	\begin{subequations}
		\label{eq:thm:predictive CBF aux 2}
		\begin{align}
			\label{seq:thm:predictive CBF aux 2_1}
			H_{T}&(x_{0}) = h(\bm{\varphi}(t_{x_{0}}^{\ast};x_{0},\bm{u}^{\ast}_{x_{0}})) - \gamma t_{x_{0}}^{\ast} \\
			\label{seq:thm:predictive CBF aux 2_2}
			&\leq \delta - \gamma (\varepsilon+1) T \\ 
			\label{seq:thm:predictive CBF aux 2_3}
			&\leq h(\bm{\varphi}(t;x_{0},\bm{u}_{e,x_{0}}^{\ast})) - \gamma t \quad \forall t\in[\vartheta^{\ast}_{x_{0}},T(\varepsilon+1)]
		\end{align}
	\end{subequations}
	where the first inequality holds by assumption for those $ x_{0} $ considered in \emph{Step~1}; for the second inequality, we note that $ \bm{\varphi}(t;x_{0},\bm{u}_{e,x_{0}}^{\ast})\in\calV $ for all $ t\geq\vartheta^{\ast}_{x_{0}} $ and thus $ \delta \leq h(\bm{\varphi}(t;x_{0},\bm{u}_{e,x_{0}}^{\ast})) $, and $ - \gamma(\varepsilon+1)T \leq -\gamma t $ for $ t\leq T(\varepsilon +1) $. 
	In particular, \eqref{eq:thm:predictive CBF aux 2} implies that
	\begin{align}
		\label{eq:thm:predictive CBF aux 2.1}
		\min_{t\in[0,\vartheta^{\ast}_{x_{0}}]} \!\!\! h(\bm{\varphi}(t;\! x_{0},\! \bm{u}^{\ast}_{x_{0}})) \!-\! \gamma t\leq \!\!\! \min_{t\in[\vartheta^{\ast}_{x_{0}}\!,(\varepsilon+ 1)T]}\!\!\!\!\! h(\bm{\varphi}(t;\!x_{0},\!\bm{u}^{\ast}_{x_{0}})) \!-\! \gamma t
	\end{align}
	where the left-hand side equals~\eqref{seq:thm:predictive CBF aux 2_1}, and the right-hand side is the minimum of~\eqref{seq:thm:predictive CBF aux 2_3}. By using the latter result, we derive
	\begin{align}
		H_{T}(x_{0}) &= \min_{t\in[0,T]} h(\bm{\varphi}(t;x_{0},\bm{u}^{\ast}_{x_{0}})) - \gamma t \nonumber\\
		\label{eq:thm:predictive CBF aux 3}
		 &\stackrel{\eqref{eq:thm:predictive CBF aux 2.1}}{=} \min_{t\in[0,(\varepsilon+1)T]} h(\bm{\varphi}(t;x_{0},\bm{u}_{e,x_{0}}^{\ast})) - \gamma t,
	\end{align}
	where the last equality follows as the extension of the time interval by $ \varepsilon T $ does not impact the value of $ H_{T} $ due to~\eqref{eq:thm:predictive CBF aux 2.1}. Based on this, and by introducing the auxiliary time variable~$ t'\in(0,\varepsilon T) $ (it may take arbitrary values on its interval of definition), we derive that
	\begin{subequations}
		\label{eq:thm:predictive CBF aux 4}
		\begin{align}
%			\label{seq:thm:predictive CBF aux 4.1}
%			&H_{T}(x_{0}) = \min_{t\in[0,T]} h(\bm{\varphi}(t;x_{0},\bm{u}_{e,x_{0}}^{\ast})) - \gamma t \\
			\label{seq:thm:predictive CBF aux 4.2}
			&H_{T}(x_{0})\stackrel{\eqref{eq:thm:predictive CBF aux 3}}{=} \min_{t\in[0,(\varepsilon+1)T]}
			h(\bm{\varphi}(t;x_{0},\bm{u}_{e,x_{0}}^{\ast})) - \gamma t \\
			\label{seq:thm:predictive CBF aux 4.3}
			&\leq \min_{t\in[\varepsilon T,(\varepsilon+1)T]} h(\bm{\varphi}(t;x_{0},\bm{u}_{e,x_{0}}^{\ast})) - \gamma t 
		\end{align}
		\begin{align}
%			\label{seq:thm:finite horizon construction main thm aux 4.4}
%			&= \min_{t\in[t',(\varepsilon+1)T]} h(\bm{\varphi}(t;x_{0},\bm{u}_{e,x_{0}}^{\ast})) - \gamma (t-t') - \gamma t' \\
			\label{seq:thm:predictive CBF aux 4.5}  
			&= \min_{t\in[0,T]} h(\bm{\varphi}(t+\varepsilon T;x_{0},\bm{u}_{e,x_{0}}^{\ast})) - \gamma t - \gamma \varepsilon T \\
			\label{seq:thm:predictive CBF aux 4.7}
			&= \min_{t\in[0,T]} h(\bm{\varphi}(t;\bm{\varphi}(\varepsilon T;x_{0},\bm{u}_{x_{0}}^{\ast}),\bm{u}_{e,x_{0}}^{\ast}(\ast\!-\!\varepsilon T))) - \gamma t - \gamma \varepsilon T \\
			\label{seq:thm:predictive CBF aux 4.8}
			&\leq \max_{\substack{\bm{u}(\cdot)\in\bm{\calU}_{[0,T]} \\ \text{s.t. } \bm{x}(0)=\bm{\varphi}(\varepsilon T;x_{0},\bm{u}^{\ast}_{x_{0}}), \\ \text{\eqref{seq:finite horizon construction H dynamics}--\eqref{seq:finite horizon construction H terminal constraint}}}} \min_{t\in[0,T]} h(\bm{x}(t)) - \gamma t - \gamma \varepsilon T \\
			\label{seq:thm:predictive CBF aux 4.9}
			&\stackrel{\eqref{eq:finite horizon construction H}}{=} H_{T}(\bm{\varphi}(\varepsilon T;x_{0},\bm{u}^{\ast}_{x_{0}})) - \gamma \varepsilon T ,
		\end{align}
	\end{subequations}
	where we obtain \eqref{seq:thm:predictive CBF aux 4.5} by employing a time-shift in the argument of the min-operator, and in~\eqref{seq:thm:predictive CBF aux 4.7} we indicate that the argument of the input trajectory is shifted by a time $ \Delta t $ by writing $ \bm{u}_{e,x_{0}}^{\ast}(\ast-\Delta t) $. We point out that the inequality in~\eqref{seq:thm:predictive CBF aux 4.8} follows from the suboptimality of $ \bm{u}_{e,x_{0}}^{\ast} $ with respect to the shifted initial state, which is $ \bm{\varphi}(\varepsilon T;x_{0},\bm{u}_{x_{0}}^{\ast}) $. To summarize the reasoning in \emph{Step~1} so far, we have shown that
	\begin{align}
		\label{eq:thm:predictive CBF aux 5}
		\frac{H_{T}(\bm{\varphi}(\varepsilon T;x_{0},\bm{u}^{\ast}_{x_{0}}))-H_{T}(x_{0})}{\varepsilon T} \geq \gamma \qquad \forall t'\in[0,T].
	\end{align}
	Furthermore, we observe that 
	\begin{subequations}
		\label{eq:thm:predictive CBF aux 5b}
		\begin{align}
			\label{seq:thm:predictive CBF aux 5b.1}
			&\liminf_{\sigma \downarrow 0} \frac{H_{T}(\bm{\varphi}(\sigma;x_{0},\bm{u}))-H_{T}(x_{0})}{\sigma} \\
			\label{seq:thm:predictive CBF aux 5b.2}
			\begin{split}
				&= \liminf_{\sigma\downarrow 0} \frac{H_{T}(x_{0}+\sigma f(x_{0},\bm{u}(0))) - H_{T}(x_{0})}{\sigma} \\
				&\quad+ \liminf_{\sigma\downarrow 0} \frac{H_{T}(\bm{\varphi}(\sigma;x_{0},\bm{u})) - H_{T}(x_{0}+\sigma f(x_{0},\bm{u}(0)))}{\sigma} 
			\end{split}
			\\
			\label{seq:thm:predictive CBF aux 5b.3}
			&= \liminf_{\sigma\downarrow 0} \frac{H_{T}(x_{0}+\sigma f(x_{0},\bm{u}(0))) - H_{T}(x_{0})}{\sigma}
		\end{align}
	\end{subequations}
	where the second limit in~\eqref{seq:thm:predictive CBF aux 5b.2} equals zero as
	\begin{align*}
		\bm{\varphi}(\sigma;x_{0},\bm{u}) &= \bm{\varphi}(0;x_{0},\bm{u}) + \sigma f(x_{0},\bm{u}(0)) + \calO(\sigma^{2}) \\
		&=x_{0} + \sigma f(x_{0},\bm{u}(0)) + \calO(\sigma^{2})
	\end{align*}
	and thus, by employing the local Lipschitz continuity of $ H_{T} $ in terms of the local Lipschitz constant $ L $,
	\begin{align*}
		&\liminf_{\sigma\downarrow 0} \frac{|H_{T}(\bm{\varphi}(\sigma;x_{0},\bm{u})) - H_{T}(x_{0}+\sigma f(x_{0},\bm{u}(0)))|}{\sigma} \\
		&\!\!=\! \liminf_{\sigma\downarrow 0} \!\frac{|H_{T}\!(x_{0} \!\!+\!\! \sigma f\!(x_{0},\!\bm{u}\!(0)) \!\!+\!\! \calO(\sigma^{2})) \!-\! H_{T}\!(x_{0}\!\!+\!\!\sigma f\!(x_{0},\!\bm{u}\!(0)))|}{\sigma} \\
		&\!\!\leq\! \liminf_{\sigma\downarrow 0} \frac{L\calO(\sigma^{2})}{\sigma} = 0.
	\end{align*}
	Equipped with these results, we can finally show that for any state $ x_{0} $ with $ H_{T}(x_{0})\leq \delta-\gamma (\varepsilon+1)T $ there exists an input trajectory, namely $ \bm{u}^{\ast}_{x_{0}} $, that results in an ascend on $ H_{T} $ in the direction of the state trajectory $ \bm{\varphi}(\cdot;x_{0},\bm{u}_{x_{0}}^{\ast}) $ at point $ x_{0} $. More precisely, it holds
	\begin{subequations}
		\label{eq:thm:predictive CBF aux 6}
		\begin{align}
			\label{seq:thm:predictive CBF aux 6.1}
			\sup_{u\in\calU}&\left\{dH_{T}(x_{0};f(x_{0},u))\right\}  \geq dH_{T}(x_{0};f(x_{0},\bm{u}_{x_{0}}^{\ast}(0))) \\
			\label{seq:thm:predictive CBF aux 6.3}
			&= \liminf_{\sigma\downarrow 0}\frac{H_{T}(x_{0}+\sigma f(x,\bm{u}_{x_{0}}^{\ast}(0))) - H_{T}(x_{0})}{\sigma} \\
			&\stackrel{\eqref{eq:thm:predictive CBF aux 5b}}{=} \liminf_{\sigma\downarrow 0}\frac{H_{T}(\bm{\varphi}(\sigma;x_{0},\bm{u}_{x_{0}}^{\ast})) - H_{T}(x_{0}))}{\sigma}\\
			\label{seq:thm:predictive CBF aux 6.4}
			&\stackrel{\eqref{eq:thm:predictive CBF aux 5}}{\geq} \lim_{\sigma\downarrow0} \frac{\gamma \sigma}{\sigma} = \gamma \\
			\label{seq:thm:predictive CBF aux 6.5}
			&\geq -\alpha(H_{T}(x_{0}))
		\end{align}
	\end{subequations}
	for all $ x_{0} $ with $ H_{T}(x_{0})\leq \delta-\gamma (\varepsilon+1)T $ where $ \alpha $ is some suitable extended class $ \calK_{e} $ function	that we choose as
	\begin{align}
		\label{eq:thm:predictive CBF aux 6.1}
		\alpha(\zeta) =
		\begin{cases}
			\alpha_{1}(\zeta) & \zeta\geq0 \\
			\alpha_{2}(\zeta) & \zeta<0
		\end{cases}
	\end{align}
	such that
	\begin{align}
		\label{eq:thm:predictive CBF aux 7}
		-\gamma \leq \alpha(\zeta)
	\end{align}
	holds. To satisfy condition~\eqref{eq:thm:predictive CBF aux 7}, we choose $ \alpha_{1}:\bbR_{\geq 0}\rightarrow\bbR_{\geq 0} $ as an arbitrary class~$ \calK $ function which trivially satisfies~\eqref{eq:thm:predictive CBF aux 7} as $ \alpha_{1} $ is by definition non-negative and $ \gamma > 0 $. In order to complement $ \alpha $ in~\eqref{eq:thm:predictive CBF aux 6.1} as an extended class~$ \calK_{e} $ function, $ \alpha_{2}:\bbR_{<0}\rightarrow\bbR_{<0} $ needs to be chosen as some strictly increasing function with $ \alpha_{2}(0)=0 $. Because also~\eqref{eq:thm:predictive CBF aux 7} needs to be satisfied, we choose $ \alpha_{2} $ as a sigmoid function but also other choices are feasible. For example, we choose $ \alpha_{2}(\zeta) = 2\gamma(\text{sig}(\zeta)-0.5) $ where $ \text{sig}(\zeta) = \frac{1}{1+e^{-\zeta}} $. A schematic sketch of $ \alpha $ can be found in Figure~\ref{fig:alpha_schematic}.
	
	\begin{figure}[t]
		\centering
		\def\svgwidth{0.7\columnwidth}
		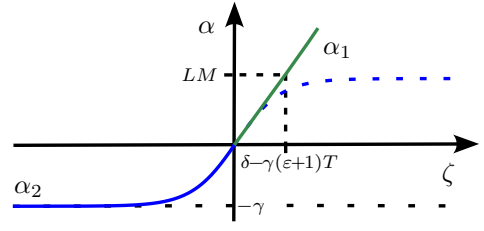
		\caption{Schematic sketch of the extended class~$ \calK_{e} $ function $ \alpha $.}
		\label{fig:alpha_schematic}
		\vspace{-\baselineskip}
	\end{figure}
	
	Now, we have shown that \eqref{eq:thm:predictive CBF aux 0} holds for all $ x_{0}\in\bbR^{n} $ with $ H_{T}(x_{0})\leq \delta-\gamma (\varepsilon+1)T $. It remains to show that~\eqref{eq:thm:predictive CBF aux 0} also holds for those $ x_{0}\in\calD $ where $ H_{T}(x_{0})\geq \delta-\gamma (\varepsilon+1)T $, which is done in the next step.

	\emph{Step~2:} Let us now consider states $ x_{0} $ with $ H_{T}(x_{0})> \delta-\gamma (\varepsilon+1)T $. By~\eqref{eq:thm:predictive CBF 0}, we then have $ H_{T}(x_{0}) > 0 $ and $ x_{0}\in\calC $. Therefore, for any such $ x_{0} $, it holds by assumption that there exists some $ u\in\calU $ such that $ ||f(x_{0},u)||\leq M $. Because $ x(t) = x_{0} + \int_{0}^{t} f(x(s),u(s)) ds $, it follows for sufficiently small $ t' $ that
	\begin{align}
		\label{eq:thm:predictive CBF aux 8}
		&||x_{0} - \bm{\varphi}(t';x_{0},\bm{u})|| = \left|\left| \int_{0}^{t'} f(\bm{\varphi}(s;x_{0},\bm{u}),\bm{u}(s)) \, ds \right|\right| \nonumber\\
		&\quad\leq \int_{0}^{t'} \!\!\! ||f(\bm{\varphi}(s;x_{0},\bm{u}),\bm{u}(s))|| \,ds \leq \int_{0}^{t'} \!\!\! M \, ds = M \, t' .
	\end{align}
	By invoking the local Lipschitz continuity of $ H_{T} $ with local Lipschitz constant $ L $, we lower-bound $ H_{T}(\bm{\varphi}(t';x_{0},\bm{u}')) $ for some input trajectory $ \bm{u}'\in\bm{\calU}_{[0,t']} $ as
	\begin{align}
		\label{eq:thm:predictive CBF aux 9}
		\begin{split}
			H_{T}(\bm{\varphi}(t';x_{0},\bm{u})) &\geq H_{T}(x_{0}) - L \, ||x_{0}-\bm{\varphi}(t';x_{0},\bm{u}')|| \\
			&\stackrel{\eqref{eq:thm:predictive CBF aux 8}}{\geq} H_{T}(x_{0}) - L\, M\, t'.
		\end{split}
	\end{align}
	Now it follows analogously to~\eqref{eq:thm:predictive CBF aux 6} that
	\begin{subequations}
		\label{eq:thm:predictive CBF aux 10}
		\begin{align}
			\label{seq:thm:predictive CBF aux 10.1}
			\sup_{u\in\calU}&\left\{dH_{T}(x_{0};f(x_{0},u))\right\} \geq dH_{T}(x_{0};f(x_{0},\bm{u}_{x_{0}}^{\ast}(0)))\\
			\label{seq:thm:predictive CBF aux 10.2}
			&= \liminf_{\sigma\downarrow 0}\frac{H_{T}(x_{0}+\sigma f(x,\bm{u}_{x_{0}}^{\ast}(0))) - H_{T}(x_{0})}{\sigma}\\
			\label{seq:thm:predictive CBF aux 10.3}
			&\stackrel{\eqref{eq:thm:predictive CBF aux 5b}}{=} \liminf_{\sigma\downarrow 0}\frac{H_{T}(\bm{\varphi}(\sigma;x_{0},\bm{u}_{x_{0}}^{\ast})) - H_{T}(x_{0}))}{\sigma}\\
			\label{seq:thm:predictive CBF aux 10.4}
			&\stackrel{\eqref{eq:thm:predictive CBF aux 9}}{\geq} \lim_{\sigma\downarrow 0} \frac{- L\, M\, \sigma}{\sigma} = -L \, M. 
		\end{align}
	\end{subequations}
	Thus, by continuously extending the class~$ \calK $ function $ \alpha_{1} $ in~\eqref{eq:thm:predictive CBF aux 6.1} such that $ \alpha_{1}(\zeta)\geq L \, M  $ for all $ \zeta> \delta-\gamma (\varepsilon+1)T $, \eqref{eq:thm:predictive CBF aux 0} also holds for all $ x_{0} $ with $ H_{T}(x_{0})> \delta-\gamma (\varepsilon+1)T $. The choice of $ \alpha_{1} $ in this step does not conflict with the choice of $ \alpha_{1} $ in Step~1 as $ \alpha_{1} $ is only required to be an arbitrary class~$ \calK $ function. 
	
	Altogether, we have now shown that~\eqref{eq:thm:predictive CBF aux 0} holds for all $ x_{0}\in\calD $. Together with the assumption that $ H_{T} $ is locally Lipschitz continuous, it follows that $ H_{T} $ is a CBF in the Dini sense according to Definition~\ref{def:cbf dini}, which concludes the proof.
\end{proof}

Conceptually, we showed in the first step of the proof that for all states marked blue in Figure~\ref{fig:predictive CBF set relation} at least an ascend of $ \gamma $ can be achieved on $ H_{T} $. In the second step, we lower-bounded the minimal possible descend on $ H_{T} $ for all states marked green in Figure~\ref{fig:predictive CBF set relation}. Based on these bounds, we established that $ H_{T} $ is a CBF in the Dini sense by constructing an extended class~$ \calK_{e} $ function satisfying~\eqref{eq:def cbf dini}. The transition between \emph{no descend} (for $ H_{T}(x_{0}) = 0 $) and \emph{limited descend} (for $ H_{T}(x_{0}) = \delta-\gamma(\varepsilon+1)T $) on $ H_{T} $ takes place on the interval $ 0\leq \zeta \leq \delta - \gamma (\varepsilon+1)T $ and is characterized by the choice of $ \alpha_{1}(\zeta) $ as a continuous function.

%At this point, some remarks are due.
%
%\begin{remark}
%	In Theorem~\ref{thm:predictive CBF}, the assumption that for any $ x\in\calC $ there exists a $ u\in\calU $ such that $ ||f(x,u)||\leq M $ is slightly too strong. It is sufficient if this assumption only holds for all $ x\in\calC_{-\delta+\gamma (\varepsilon+1)T} = \{x \, | \, H_{T}(x)\geq \delta-\gamma (\varepsilon+1)T\} $. This is because the assumption is only required in step~2 of the proof.
%\end{remark}

\begin{remark}
	The assumption of Theorem~\ref{thm:predictive CBF} that $ f $ is bounded on $ \calC $ is reasonable as the time-variation of practically relevant systems is bounded within their domain of operation. 
\end{remark}

\begin{remark}
	It is known for certain types of nonlinear systems that finite horizon optimal control problems such as~\eqref{eq:finite horizon construction H} may be sensitive to variations in the initial condition. While this seems to be a minor practical problem, it can be still observed in some examples~\cite{Grimm2004}. This can be mitigated by varying the design parameters $ T $, $ \gamma $, $ \delta $ or choosing $ \calF $ closer to $ \calV $. In Theorem~\ref{thm:predictive CBF}, we account for this by including local Lipschitz continuity into the premises. This can be observed after the computation of $ H_{T} $.
\end{remark}

\subsection{Relaxed Upper-Bound on $ \gamma $}

The proof of the previous theorem unraveled the role of parameter~$ \gamma $ in the CBF synthesis. As it becomes evident from~\eqref{seq:thm:predictive CBF aux 6.4}, $ \gamma $ constitutes a lower bound on the maximum possible ascend on $ H_{T} $ at any $ x_{0} $ contained in the set marked blue in Figure~\ref{fig:predictive CBF set relation}. In Theorem~\ref{thm:predictive CBF}, we formulated a condition on the choice of $ \gamma $ in~\eqref{eq:thm:predictive CBF 0}. It can be relaxed as follows: 

\begin{itemize}
	\item If  
	\begin{align}
		\label{eq:minimization horizon}
		t^{\ast}_{x_{0}} \coloneq \argmin_{t\in[0,T]} h(\bm{\varphi}(t; x_{0},\bm{u}_{x_{0}}^{\ast})) - \gamma t \in [0,\overline{T}]
	\end{align}
	for some time $ \overline{T} < T $ and any $ x_{0} $, where $ t^{\ast}_{x_{0}} $ is defined as the time that solves the inner minimization in~\eqref{seq:finite horizon construction H max min}, then condition~\eqref{eq:thm:predictive CBF 0} relaxes to $ \gamma < \frac{\delta}{\overline{T}} $.
	\item Or, introducing the extended class~$ \calK_{e} $ function $ \alpha $ into~\eqref{eq:finite horizon construction H} instead of $ \gamma $ allows us to drop condition~\eqref{eq:thm:predictive CBF 0}.
\end{itemize}

While the first approach requires additional knowledge, the second adds slightly to the complexity of the computation of~$ H_{T} $. We investigate both approaches in the following.

\begin{theorem}
	\label{cor:gamma condition refined}
	Let~\eqref{eq:minimization horizon} hold for some time $ \overline{T} < T $. Then Theorem~\ref{thm:predictive CBF} holds with
	\begin{align}
		\label{eq:cor:gamma condition refined 0}
		\gamma < \frac{\delta}{\overline{T}}
	\end{align}
	instead of condition~\eqref{eq:thm:predictive CBF 0}.
\end{theorem}
\begin{proof}
	Let us consider $ \varepsilon \in (0,\min\{\frac{\delta}{\gamma \overline{T}}, \frac{T}{\overline{T}}\}-1) $. Then, $ \gamma < \frac{\delta}{(\varepsilon+1)\overline{T}} < \frac{\delta}{\overline{T}} $ and $ (\varepsilon+1)\overline{T} < T $ hold. Analogously to before, we divide the proof into two steps. 
	
	\emph{Step~1:} Let $ x_{0} $ be such that  $ H_{T}(x_{0})\leq \delta-\gamma (\varepsilon+1)\overline{T} $. We note that the right-hand side is strictly positive due to the above choice of $ \varepsilon $. Furthermore, let us consider $ \bm{u}^{\ast}_{e,x_{0}} $ as defined in~\eqref{eq:thm:predictive CBF aux 1}. Based on~\eqref{eq:minimization horizon}, we derive 
	\begin{subequations}
		\label{eq:cor:gamma condition refined 1}
		\begin{align}
			\label{seq:cor:gamma condition refined 1_1}
			H_{T}(x_{0}) &= \min_{t\in[0,T]} h(\bm{\varphi}(t;x_{0},\bm{u}_{e,x_{0}}^{\ast})) - \gamma t \\
			\label{seq:cor:gamma condition refined 1_2}
			& \stackrel{\eqref{eq:minimization horizon}}{=} \min_{t\in[0,\overline{T}]} h(\bm{\varphi}(t;x_{0},\bm{u}_{e,x_{0}}^{\ast})) - \gamma t \\
			\label{seq:cor:gamma condition refined 1_3}
			&\stackrel{\eqref{eq:minimization horizon}}{=} \min_{t\in[0,(\varepsilon+1)\overline{T}]} h(\bm{\varphi}(t;x_{0},\bm{u}_{e,x_{0}}^{\ast})) - \gamma t \\
			\label{seq:cor:gamma condition refined 1_4}
			&\leq \min_{t\in[\varepsilon T,(\varepsilon+1)\overline{T}]} h(\bm{\varphi}(t;x_{0},\bm{u}_{e,x_{0}}^{\ast})) - \gamma t\\
			\label{seq:cor:gamma condition refined 1_5}
			&\stackrel{\eqref{eq:minimization horizon}}{=} \min_{t\in[\varepsilon T,(\varepsilon+1)T]} h(\bm{\varphi}(t;x_{0},\bm{u}_{e,x_{0}}^{\ast})) - \gamma t
		\end{align} 
	\end{subequations}
	where \eqref{seq:cor:gamma condition refined 1_2} and \eqref{seq:cor:gamma condition refined 1_5} are immediate consequences of~\eqref{eq:minimization horizon}; in~\eqref{seq:cor:gamma condition refined 1_3}, we additionally employed that for the chosen $ \varepsilon $ it holds $ (\varepsilon+1)\overline{T} < T $. We note that \eqref{seq:cor:gamma condition refined 1_5} coincides with~\eqref{seq:thm:predictive CBF aux 4.3} in the proof of Theorem~\ref{thm:predictive CBF}, and we follow the same steps from here onward. Thereby, \eqref{eq:thm:predictive CBF aux 0} follows for all $ x_{0} $ considered in Step~1. 
	
	\emph{Step~2:} We follow the proof of Theorem~\ref{thm:predictive CBF}, Step~2, for all $ x_{0} $ with $ H_{T}(x_{0})> \delta-\gamma (\varepsilon+1)\overline{T} $ and finally choose $ \alpha_{1} $ such that $ \alpha_{1}(\zeta)\geq L\,M $ for all $ \zeta > \delta - \gamma(\varepsilon+1)\overline{T} $.
\end{proof}

\begin{remark}
	The construction of~$ \alpha $ in the proof of Theorem~\ref{cor:gamma condition refined} stays the same as in Figure~\ref{fig:alpha_schematic}, however, with $ T $ replaced by~$ \overline{T} $.
\end{remark}

Let us assume a differentiable constraint function~$ h $. Then according to the KKT-conditions, it holds at time~$ t^{\ast}_{x_{0}} $ in~\eqref{eq:minimization horizon} that $ \nabla h(\bm{x}(t^{\ast}_{x_{0}})) \, f(\bm{x}(t^{\ast}_{x_{0}}),u) \geq \gamma $ and $ \frac{d}{dt} (\nabla h(\bm{x}(t^{\ast}_{x_{0}})) \, f(\bm{x}(t^{\ast}_{x_{0}}),u)) >0 $ for some $ u\in\calU $. We apply this insight in the next example.

\noindent
\parbox{0.4\columnwidth}{%
	\centering
	\def\svgwidth{0.3\columnwidth}
	%% Creator: Inkscape 1.3.2 (091e20e, 2023-11-25, custom), www.inkscape.org
%% Accompanies image file 'optimized_horizon_bicycle.pdf' (pdf, eps, ps)
\begingroup%
  \makeatletter%
  \providecommand\color[2][]{%
    \errmessage{(Inkscape) Color is used for the text in Inkscape, but the package 'color.sty' is not loaded}%
    \renewcommand\color[2][]{}%
  }%
  \providecommand\transparent[1]{%
    \errmessage{(Inkscape) Transparency is used (non-zero) for the text in Inkscape, but the package 'transparent.sty' is not loaded}%
    \renewcommand\transparent[1]{}%
  }%
  \providecommand\rotatebox[2]{#2}%
  \newcommand*\fsize{\dimexpr\f@size pt\relax}%
  \newcommand*\lineheight[1]{\fontsize{\fsize}{#1\fsize}\selectfont}%
  \ifx\svgwidth\undefined%
    \setlength{\unitlength}{249.59069824bp}%
    \ifx\svgscale\undefined%
      \relax%
    \else%
      \setlength{\unitlength}{\unitlength * \real{\svgscale}}%
    \fi%
  \else%
    \setlength{\unitlength}{\svgwidth}%
  \fi%
  \global\let\svgwidth\undefined%
  \global\let\svgscale\undefined%
  \makeatother%
  \begin{picture}(1,0.82654109)%
    \lineheight{1}%
    \setlength\tabcolsep{0pt}%
    \put(0,0){\includegraphics[width=\unitlength,page=1]{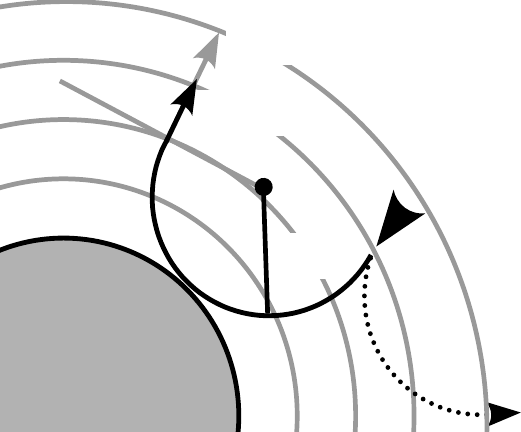}}%
    \put(0.53611976,0.31045209){\color[rgb]{0,0,0}\makebox(0,0)[lt]{\lineheight{1.25}\smash{\begin{tabular}[t]{l}\footnotesize{$R$}\end{tabular}}}}%
    \put(0.38384609,0.57707333){\color[rgb]{0,0,0}\makebox(0,0)[lt]{\lineheight{1.25}\smash{\begin{tabular}[t]{l}\footnotesize{$f(x,u)$}\end{tabular}}}}%
    \put(0.43588403,0.71906395){\color[rgb]{0,0,0}\makebox(0,0)[lt]{\lineheight{1.25}\smash{\begin{tabular}[t]{l}\footnotesize{$\nabla h(x)$}\end{tabular}}}}%
  \end{picture}%
\endgroup%

	\captionof{figure}{}
	\label{fig:optimized_horizon_bicycle}
}
\hfill
\parbox{0.59\columnwidth}{%
	\begin{example}
		Let us reconsider the bicycle dynamics in Example~\ref{exmp:bicycle1}. Furthermore, let $ h $ be a differentiable function with $ ||\nabla h(x)|| = 1 $ specifying some convex constraint. Now we note that the bicycle can be steered from any initial point $ x_{0} $ to a point~$ x $ 
	\end{example}
	\vspace{-\parskip}
}
\noindent where $ \nabla h(x) $ and $ f(x,u) $ are aligned as depicted in Figure~\ref{fig:optimized_horizon_bicycle}. This can be done by steering the bicycle on a circular trajectory with radius~$ R $ for at most a half round; $ R $ denotes the turning radius as previously specified. Thus if $ v_{\text{max}} \geq \gamma $, we obtain that $ \nabla h(x) \, f(x,u) \geq ||\nabla h(x)|| \, ||f(x,u)|| \geq v_{\text{max}} \geq \gamma $ and hence that the KKT conditions are satisfied for some $ t^{\ast}_{x_{0}}\leq\overline{T} = \frac{\pi R}{v_{\text{max}}} $.

\subsection{Incorporating $ \alpha $ as Explicit Design Parameter}

Alternatively, the extended class $ \calK_{e} $ function $ \alpha $ can be used as a design parameter in the synthesis of $ H_{T} $  instead of $ \gamma $. To this end, we modify~\eqref{eq:finite horizon construction H} and replace $ \gamma $ by 
\begin{align}
	\label{eq:alpha_bar}
	\bar{\alpha}(\zeta) \coloneq \begin{cases} \alpha(\zeta) & \text{if } \zeta\leq 0 \\ 0 & \text{if } \zeta>0 \end{cases}
\end{align}
where $ \alpha $ is an extended class $ \calK_{e} $ function. The modified optimization problem is then
\begin{subequations}
	\label{eq:finite horizon construction H alpha}
	\begin{align}
		\label{seq:finite horizon construction H max min alpha}
		H_{T}(x_{0}) &\coloneq \max_{\bm{u}(\cdot)\in\bm{\calU}_{[0,T]}} \min_{t\in[0,T]} h(\bm{x}(t)) + \bar{\alpha}(h(\bm{x}(t))) \, t \\
		\label{seq:finite horizon construction H initial condition alpha}
		\text{s.t.}\;\; & \text{\eqref{seq:finite horizon construction H initial condition}-\eqref{seq:finite horizon construction H terminal constraint} hold}.
	\end{align}
\end{subequations}
By following similar arguments as before, we can show that also $ H_{T} $ as defined in~\eqref{eq:finite horizon construction H alpha} constitutes a CBF in the Dini sense. 

\begin{theorem}
	\label{thm:predictive CBF alpha}
	Let Assumptions~\ref{ass:setF} and~\ref{ass:controllability} hold, and let $ h $ and $ T $ be as in Theorem~\ref{thm:predictive CBF}. Moreover, let $ H_{T}:\calD \rightarrow \bbR $ be defined on some domain $ \calD\subseteq\bbR^{n} $ with $ \calH \subset \calD $ in~\eqref{eq:finite horizon construction H alpha} where $ \bar{\alpha} $ is defined as in \eqref{eq:alpha_bar} via some extended class $ \calK_{e} $ function $ \alpha $. Furthermore, let $ f $ be bounded in the sense that for all $ x $ with $ H_{T}(x)\geq\delta $ there exists a $ u\in\calU $ such that $ ||f(x,u)||\leq M $ for some constant $ M>0 $. If $ H_{T} $ is locally Lipschitz continuous with Lipschitz constant $ L $, and $ \alpha $ satisfies $ \alpha(\zeta) \geq L\,M $ for all $ \zeta \geq \delta $, then 
	\begin{align}
		\label{eq:thm:predictive CBF alpha}
		\sup_{u\in\calU}\left\{dH_{T}(x_{0};f(x_{0},u)) \right\} \geq -\alpha(H_{T}(x_{0}))
	\end{align}
	and $ H_{T} $ constitutes a CBF in the Dini sense on the domain $ \calD $ with respect to the dynamics~\eqref{eq:dynamics}.
\end{theorem}
\begin{proof}
	Using the same arguments as in Proposition~\ref{prop:welldefined H_T and C subseteq H}, we conclude that also $ H_{T} $ defined in~\eqref{eq:finite horizon construction H alpha} is well-defined for all $ x_{0}\in\calD $. We conduct this proof, as the previous ones, in two steps: at first, we consider those $ x_{0} $ with $ H_{T}(x_{0}) \leq \delta $; in a second step, $ x_{0} $ with $ H_{T}(x_{0})> \delta $ are considered. In each of the steps, we mostly follow the line of arguments in the proof of Theorem~\ref{thm:predictive CBF} and therefore only point out important intermediate results. In this proof, we denote the input trajectory $ \bm{u} $ and the times $ t $ and $ \vartheta $ that solve optimization problem~\eqref{eq:finite horizon construction H alpha} by $ \bm{u}^{\ast}_{x_{0}} $, $ t^{\ast}_{x_{0}} $ and $ \vartheta_{x_{0}}^{\ast} $ analogously to before.
	
	\emph{Step~1:} Let $ x_{0} $ be such that $ H_{T}(x_{0})\leq\delta $. As in the proof of Theorem~\ref{thm:predictive CBF}, we obtain analogously to~\eqref{eq:thm:predictive CBF aux 2}
	\begin{align*}
		H_{T}(x_{0}) &= h(\bm{\varphi}(t_{x_{0}}^{\ast};x_{0},\bm{u}^{\ast}_{x_{0}})) + \bar{\alpha}(h(\bm{\varphi}(t_{x_{0}}^{\ast};x_{0},\bm{u}^{\ast}_{x_{0}}))) \, t_{x_{0}}^{\ast} \nonumber \\
%		\label{eq:thm:predictive CBF alpha aux 2}
		&\leq \delta \leq h(\bm{\varphi}(t;x_{0},\bm{u}_{e,x_{0}}^{\ast})) \qquad \forall t\in[\vartheta^{\ast}_{x_{0}},T(\varepsilon\!+\!1)]
	\end{align*}
	where $ \bm{u}_{e,x_{0}}^{\ast} \coloneq \left\{\begin{smallmatrix*}[l] \bm{u}^{\ast}_{x_{0}}(t) &\text{if } t\in[0,\vartheta^{\ast}_{x_{0}}] \\
	\bm{u}_{e,x_{0}}(t) &\text{if } t\in(\vartheta^{\ast}_{x_{0}},T(\varepsilon+1)] \end{smallmatrix*}\right. $ with $ \bm{u}_{e,x_{0}} \in \bm{\calU}_{(\vartheta^{\ast}_{x_{0}},T(\varepsilon+1)]} $ such that $ \bm{\varphi}(t;x_{0},\bm{u}^{\ast}_{e,x_{0}})\in\calV $ for all $ t>\vartheta^{\ast}_{x_{0}} $. Thus the analogue to~\eqref{eq:thm:predictive CBF aux 3} becomes 
	\begin{align*}
		&H_{T}(x_0) = \min_{t\in[0,T]} h(\bm{\varphi}(t;x_{0},\bm{u}^{\ast}_{x_{0}})) + \bar{\alpha}(h(\bm{\varphi}(t;x_{0},\bm{u}^{\ast}_{x_{0}}))) \, t \nonumber\\
%		\label{eq:thm:predictive CBF alpha aux 3}
		&\quad\leq \min_{t\in[0,T(\varepsilon+1)]} h(\bm{\varphi}(t;x_{0},\bm{u}^{\ast}_{e,x_{0}})) + \bar{\alpha}(h(\bm{\varphi}(t;x_{0},\bm{u}^{\ast}_{e,x_{0}}))) \, t
	\end{align*}
	which is the same as \eqref{eq:thm:predictive CBF aux 3} when replacing $ \gamma $ with $-\bar{\alpha}$; note that $-\bar{\alpha}$ is by definition non-negative. This ultimately leads to 
	\begin{align}
		\label{eq:thm:predictive CBF alpha aux 5}
		\frac{H_{T}(\bm{\varphi}(\varepsilon;x_{0},\bm{u}^{\ast}_{x_{0}}))-H_{T}(x_{0})}{\varepsilon} \geq -\bar{\alpha}(H_{T}(\bm{\varphi}(\varepsilon;x_{0},\bm{u}^{\ast}_{x_{0}})))
	\end{align}
	for all $ \varepsilon\in[0,T] $, and further to
	\begin{subequations}
		\label{eq:thm:predictive CBF alpha aux 6}
		\begin{align}
			\label{seq:thm:predictive CBF alpha aux 6.1}
			&\sup_{u\in\calU}\left\{dH_{T}(x_{0};f(x_{0},u))\right\}  \geq dH_{T}(x_{0};f(x_{0},\bm{u}_{x_{0}}^{\ast}(0))) \\
			\label{seq:thm:predictive CBF alpha aux 6.3}
			&\quad= \liminf_{\varepsilon\downarrow 0}\frac{H_{T}(x_{0}+\sigma f(x,\bm{u}_{x_{0}}^{\ast}(0))) - H_{T}(x_{0})}{\sigma}\\
			&\quad\stackrel{\eqref{eq:thm:predictive CBF aux 5b}}{=} \liminf_{\sigma\downarrow 0}\frac{H_{T}(\bm{\varphi}(\sigma;x_{0},\bm{u}_{x_{0}}^{\ast})) - H_{T}(x_{0}))}{\sigma}\\
			\label{seq:thm:predictive CBF alpha aux 6.4}
			&\quad\stackrel{\eqref{eq:thm:predictive CBF alpha aux 5}}{\geq} \lim_{\sigma\downarrow0} -\frac{\bar{\alpha}(H_{T}(\bm{\varphi}(\!\sigma;x_{0},\bm{u}^{\ast}_{x_{0}}\!))) \,  \sigma}{\sigma} = -\bar{\alpha}(H_{T}(x_{0})) \\
			\label{seq:thm:predictive CBF alpha aux 6.5}
			&\quad\geq -\alpha(H_{T}(x_{0})).
		\end{align}
	\end{subequations}

	\emph{Step~2:} Replacing $ \gamma $ with $ -\bar{\alpha} $ does not change Step~2 in the proof of Theorem~\ref{thm:predictive CBF}. Thus, we analogously conclude  
	\begin{align}
		\sup_{u\in\calU}&\left\{dH_{T}(x_{0};f(x_{0},u))\right\} \geq -L\,M\geq-\alpha(H_{T}(x_{0})).
	\end{align}

	Combining Step~1 and~2, it follows that \eqref{eq:thm:predictive CBF alpha} holds and $ H_{T} $ defined by~\eqref{eq:finite horizon construction H alpha} is a CBF in the Dini sense. 
	%This concludes the proof.	
\end{proof}

In the remainder of the paper, we refer to~\eqref{eq:finite horizon construction H} whenever we write~$ H_{T} $ unless otherwise specified. Nevertheless, results analogous to the following ones also apply to $ H_{T} $ in~\eqref{eq:finite horizon construction H alpha}.

%#######################################################################################

\section{Properties of $ H_{T} $}
\label{sec:properties}

Next, we characterize some of the properties of the determined CBF~$ H_{T} $.

\subsection{Impact of the Prediction Horizon}
At first, we investigate the impact of the prediction horizon~$ T $ on the size of the zero-superlevel set~$ \calC $ as well as on the size of the other superlevel sets $ \calC_{\lambda} $ of $ H_{T} $ with $ \lambda\geq0 $. To avoid ambiguities in the notation, we include the time horizon as an argument to these sets. In particular, we define $ \calC_{0,T} \coloneq \{ x \, | \, H_{T}(x) \geq 0 \} $ and $ \calC_{\lambda,T} \coloneq \{ x \, | \, H_{T}(x) \geq -\lambda \}. $
%\begin{align*}
%	\calC_{0,T} &\coloneq \{ x \, | \, H_{T}(x) \geq 0 \},  \\
%	\calC_{\lambda,T} &\coloneq \{ x \, | \, H_{T}(x) \geq -\lambda \}.	
%\end{align*}
We show next that the sets $ \calC_{0,T} $ and $ \calC_{\lambda,T} $, $ \lambda\geq0 $, can be enlarged by increasing time horizon $ T $. 

\begin{proposition}
	\label{prop:finite horizon CBF construction impact time horizon}
	Let Assumptions~\ref{ass:setF} and~\ref{ass:controllability} hold, let $ h $ be Lipschitz continuous and let $ T_{2}\geq T_{1}\geq\tau $. Moreover, let $ H_{T_{1}}: \calD\rightarrow\bbR $ and $ H_{T_{2}}: \calD\rightarrow\bbR $ be defined as in~\eqref{eq:finite horizon construction H} for some $ \gamma>0 $ where
	\begin{align}
		\label{eq:prop:finite horizon CBF construction impact time horizon 0}
		\gamma < \frac{\delta}{T_{2}}.
	\end{align}
	Then $ \calC_{0,T_{1}} \subseteq \calC_{0,T_{2}} $ and, more generally, $ \calC_{\lambda,T_{1}} \subseteq \calC_{\lambda,T_{2}} $ for any $ \lambda\geq0 $ with $ \calC_{\lambda,T_{1}}, \calC_{\lambda,T_{2}} \subseteq\calD $. 
\end{proposition}
\begin{proof}
	The proof can be found in the appendix. 
\end{proof}

Analogous results hold for $ \gamma < \frac{\delta}{\max\{\overline{T}_{1},\overline{T}_{2}\}} $, or if $ H_{T} $ is synthesized based on~\eqref{eq:finite horizon construction H alpha}. 

\subsection{Changing and Time-Varying Constraint Specifications}

Independent of whether $ H_{T} $ was synthesized based on Theorem~\ref{thm:predictive CBF}, \ref{cor:gamma condition refined} or~\ref{thm:predictive CBF alpha}, further CBFs can be derived based on it. In particular, the CBF property of $ H_{T} $ is preserved when adding -- within bounds still to be further specified -- a constant or a time-varying trajectory. In this way, the synthesized CBF~$ H_{T} $ can account for corresponding changes in the constraint specifications and time-variations. 

To become more specific, let us define the largest superlevel set of $ H_{T} $ contained in $ \calD $ as $ \calC_{\Lambda,T} \coloneq \{ x \, | \, H_{T}(x) \geq -\Lambda \} $ where $ \Lambda \coloneq \max\{\lambda \,|\, \calC_{\lambda,T}\subseteq \calD\} $. Based on this, 
\begin{align*}
	H_{\lambda,T}(x) \coloneq H_{T}(x) + \lambda
\end{align*}
is also a CBF for any $ \lambda\in[0,\Lambda] $. This is a direct implication of the fact that\footnote{In the case of $ H_{T} $ defined by~\eqref{eq:finite horizon construction H alpha}, it holds analogously $ \sup_{u\in\calU}\left\{dH_{T}(x_{0};f(x_{0},u))\right\} \geq -\bar{\alpha}(H_{T}(x_{0})) \geq 0 $ according to~\eqref{eq:thm:predictive CBF alpha aux 6} for any~$ x_{0} $ with $ H_{T}(x_{0})\in [-\Lambda,0] $.} $ \sup_{u\in\calU}\left\{dH_{T}(x_{0};f(x_{0},u))\right\}  \geq \gamma \geq 0 $
%\begin{align*}
%	\sup_{u\in\calU}&\left\{dH_{T}(x_{0};f(x_{0},u))\right\}  \geq \lambda \geq 0
%\end{align*}
in any $ x_{0} $ where $ H_{T}(x_{0})\in [-\Lambda,0] $ as shown in \eqref{eq:thm:predictive CBF aux 6} as part of the proof of Theorem~\ref{thm:predictive CBF}. In the light of~\cite{Wiltz2024a}, an even stronger property holds. Assuming that $ H_{T} $ is not only locally Lipschitz continuous but differentiable, then $ H_{T} $ constitutes a so-called $ \Lambda $-shiftable CBF on the domain $ \calD $ with respect to~\eqref{eq:dynamics}; for details refer to~\cite{Wiltz2024a}. This characteristic leads to the following property: for a time-varying trajectory $ \bm{\lambda}: \bbR_{\geq0} \rightarrow [0,\Lambda] $ satisfying some additional condition, the function
\begin{align}
	\label{eq:H_T shifted in time}
	H_{\bm{\lambda}(\cdot),T}(t,x) \coloneq H_{T}(x) + \bm{\lambda}(t)
\end{align} 
constitutes a differentiable CBF~\cite{Ames2017} with respect to the dynamics augmented in time. This is formally stated as follows.

\begin{proposition}
	\label{prop:lambda shiftable}
	Let $ H_{\bm{\lambda}(\cdot),T} $ be defined as in~\eqref{eq:H_T shifted in time}, and let the same premises hold as in Theorem~\ref{thm:predictive CBF}. Moreover, let $ H_{T} $ as defined in~\eqref{eq:finite horizon construction H} be differentiable and let $ \bm{\lambda}: \bbR_{\geq0}\rightarrow[0,\Lambda] $ be a differentiable function that satisfies 
	\begin{align}
		\label{eq:prop:lambda shiftable}
		\frac{\partial \bm{\lambda}}{\partial t} (t) \geq -\widetilde{\alpha}(\bm{\lambda}(t))
	\end{align}
	where $ \widetilde{\alpha} $ is an either convex or concave class~$ \calK $ function and it holds $ \alpha(-\zeta) \leq -\widetilde{\alpha}(\zeta) $ for all $ \zeta\in[0,\Lambda] $. Then, $ H_{\bm{\lambda}(\cdot),T} $ is a CBF on the domain $ \bbR_{\geq0} \times \calD $ with respect to dynamics~\eqref{eq:dynamics} augmented by time $ \left[\begin{smallmatrix} \dot{t} \\ \dot{x} \end{smallmatrix}\right] = \left[\begin{smallmatrix} 1 \\ f(x,u) \end{smallmatrix}\right] $. 
\end{proposition}

\begin{proof}
	By Theorem~\ref{thm:predictive CBF}, $ H_{T} $ is a CBF on the domain $ \calC_{\Lambda}\subseteq\calD $. Hence, as $ H_{T} $ is assumed to be differentiable, it follows by \cite[Definition~2]{Wiltz2024a} that $ H_{T} $ is a $ \Lambda $-shiftable CBF. Based on this, the proposition is a direct implication of~\cite[Theorem~4]{Wiltz2024a}.
\end{proof}

\begin{remark}
	The same result follows if the premises of Theorem~\ref{cor:gamma condition refined} hold instead of those of Theorem~\ref{thm:predictive CBF}, or alternatively, if $ H_{T} $ is defined in~\eqref{eq:finite horizon construction H alpha} with the premises of Theorem~\ref{thm:predictive CBF alpha} satisfied.
\end{remark}

The construction of $ H_{T} $ in the previous section has the following favorable properties with respect to the application of Proposition~\ref{prop:lambda shiftable}: 

1.) While most CBF construction approaches in the literature only apply to the determination of CBFs on a domain~$ \calC $, our proposed method allows for the construction on larger domains $ \calD\supseteq\calC $, which gives rise to the ``shiftability''-property. 

2.) Our method comes along with a concrete construction of the extended class~$ \calK_{e} $ function $ \alpha $ invoking the design parameter $ \gamma $ as outlined in the proofs of Theorems~\ref{thm:predictive CBF} and~\ref{cor:gamma condition refined}, see Figure~\ref{fig:alpha_schematic}. Alternatively, in the synthesis of $ H_{T} $ according to~\eqref{eq:finite horizon construction H alpha}, $ \alpha $ can be even chosen as a design parameter. 

3.) Regardless of the particular approach, function~$ \alpha $ can here always be chosen to be convex (see also Figure~\ref{fig:alpha_schematic}), thus $ \alpha(-\zeta) = -\tilde{\alpha}(\zeta) $ for all $ \zeta\in[0,\Lambda] $ in Proposition~\ref{prop:lambda shiftable} is a feasible choice. Then, \eqref{eq:prop:lambda shiftable} becomes 
\begin{align}
	\label{eq:lambda condition simplified}
	\frac{\partial \bm{\lambda}}{\partial t} (t) \geq \alpha(-\bm{\lambda}(t)).
\end{align}

4.) Our CBF construction approach accounts for input constraints. 

\subsection{Saturated $ H_{T} $}\label{subsubsec:saturated H_T} By introducing a saturated version of $ H_{T} $, the computation of a CBF in the Dini sense can be simplified. To this end, we make the following observation.

\begin{proposition}
	\label{cor:saturated cbf}
	Let the saturated value function $ \widebar{H}_{T}: \calD \rightarrow \bbR $ be defined as 
	\begin{align*}
		\widebar{H}_{T}(x_{0}) \coloneq \min\{H_{T}(x_{0}),\delta - \gamma T\}
	\end{align*}
	where $ H_{T} $ is defined by~\eqref{eq:finite horizon construction H}, and let the same premises hold as in Theorem~\ref{thm:predictive CBF}. Then, $ \widebar{H}_{T} $ is a CBF in the Dini sense.
\end{proposition}
\begin{proof}
	For $ x_{0} $ with $ \widebar{H}_{T}(x_{0}) \leq \delta - \gamma T $, we trivially have $ \widebar{H}_{T}(x_{0}) = H_{T}(x_{0}) $ and the proof of Theorem~\ref{thm:predictive CBF} still applies. For $ x_{0} $ with $ \widebar{H}_{T}(x_{0}) > \delta - \gamma T $, it holds $ d\widebar{H}_{T}(x_{0};f(x_{0},u)) = 0 $ for any $ u\in\calU $ as $ \widebar{H}_{T} $ is constant in the $ \epsilon $-neighborhood of $ x $. Thus, \eqref{eq:thm:predictive CBF aux 10} still holds when substituting $ H_{T} $ by $ \widebar{H}_{T} $ because $ \sup_{u\in\calU}\left\{d\widebar{H}_{T}(x_{0};f(x_{0},u))\right\} = 0 \geq -LM $. From this, we conclude that the proof of Theorem~\ref{thm:predictive CBF} also applies to $ \widebar{H}_{T} $ and we conclude that $ \widebar{H}_{T} $ is a CBF in the Dini sense.
\end{proof}

\begin{remark}
	If the premises of Theorem~\ref{cor:gamma condition refined} are considered instead of those in Theorem~\ref{thm:predictive CBF}, then the above results holds for $ \widebar{H}_{T}(x_{0}) \coloneq \min\{H_{T}(x_{0}),\delta - \gamma \overline{T}\} $. If $ H_{T} $ as defined by~\eqref{eq:finite horizon construction H alpha} is considered and the premises of Theorem~\ref{thm:predictive CBF alpha} hold, then the above result holds for $ \widebar{H}_{T}(x_{0}) \coloneq \min\{H_{T}(x_{0}),\delta\} $. The proofs in each of the cases are analogous to the one of Proposition~\ref{cor:saturated cbf}.
\end{remark}

\section{Implementation Remarks}
\label{sec:implementation remarks}

The implementation of the optimization problems in~\eqref{eq:finite horizon construction H} and~\eqref{eq:finite horizon construction H alpha} as a max-min-problem is not entirely straightforward and deserves a discussion on its own. While their formulation is well-suited for analysis, we propose in the sequel some simplifications with regard to their practical implementation. These yield an arbitrarily close approximation of $ H_{T}(x_{0}) $ while giving rise to a computationally efficient implementation. We focus on $ H_{T} $ defined in~\eqref{eq:finite horizon construction H}, and note that analogous remarks apply to $ H_{T} $ defined in~\eqref{eq:finite horizon construction H alpha}.

\subsection{Discrete-Time Implementation} 

For the practical implementation of the optimization problems in~\eqref{eq:finite horizon construction H} and~\eqref{eq:finite horizon construction H alpha}, dynamics and trajectories need to be discretized. To this end, we discretize time-horizon $ T $ into $ N+1 $ time-steps using discretization time $ \Delta t = T/N $. Correspondingly, state and input trajectories in~\eqref{eq:finite horizon construction H} become 
\begin{align*}
	\bm{\mathsf{x}}_{N}&\!\!:=\!\!
	\begin{bmatrix}
		\bm{x}(0  \Delta\! T)\!\!\!\!\! & \bm{x}(1 \Delta\! T)\!\!\!\!\! & \bm{x}(2 \Delta\! T)\!\!\!\!\! & \dots\!\!\!\!\! & \bm{x}(N \Delta \!T)
	\end{bmatrix}
	\!\in\!\bbR^{n,N\!+\!1}\!, \\
	\bm{\mathsf{u}}_{N\!-\!1}&\!\!:=\!\!
	\begin{bmatrix}
		\bm{u}(0 \Delta\! T)\!\!\!\!\! & \bm{u}(1 \Delta\! T)\!\!\!\!\! & \bm{u}(2 \Delta\! T)\!\!\!\!\! & \dots\!\!\!\!\! & \bm{u}((N\!\!-\!\!1) \Delta\! T)
	\end{bmatrix}\!\in\!\bbR^{m,N}\!.
\end{align*}
The $ k $-th column of $ \bm{\mathsf{x}}_{N} $ is referred to as $ \bm{\mathsf{x}}_{N}[k] $. The values of~$ h $ corresponding to 
%the discretized state trajectory 
$ \bm{\mathsf{x}}_{N}[\cdot] $ are
\begin{align*}
	\bm{\mathsf{h}}_{\!N}\!(\bm{\mathsf{x}}_{\!N})\!\!:\!= \!\!
	\begin{bmatrix}
		h(\bm{\mathsf{x}}_{\! N}[0])\!\!\!\!\! & h(\bm{\mathsf{x}}_{\! N}[1])\!\!\!\!\! & h(\bm{\mathsf{x}}_{\! N}[2])\!\!\!\!\! & \dots\!\!\!\!\! & h(\bm{\mathsf{x}}_{\! N}[N]) \!
	\end{bmatrix}^{T}
	\!\!\!\in\! \bbR^{\!N\!+\!1}\!.
\end{align*}
The max-min-problem~\eqref{eq:finite horizon construction H} can be then rewritten as
\begin{subequations}
	\label{eq:H_disc}
	\begin{align}
		\label{seq:H max min_disc}
		H_{T}(x_{0}) &:= \max_{\bm{\mathsf{u}}_{N\!-\!1}} \min_{k=0,1,\dots,N} h(\bm{\mathsf{x}}_{N}[k]) - \gamma k \Delta t \\
		\label{seq:H initial condition_disc}
		\text{s.t.}\;\; &\bm{\mathsf{x}}_{\! N}[0]=x_{0},\\
		\label{seq:H dynamics_disc}
		&\bm{\mathsf{x}}_{\! N}\![k\!\!+\!\!1] \!\!=\!\! f_{\!d}(\bm{\mathsf{x}}_{\! N}\![k],\!\bm{\mathsf{u}}_{\! N\!-\!1}\![k]), \;\forall \!k\!\!=\!\!0,\!1,\dots,\!N\!\!-\!\!1,\! \\
		\label{seq:H input constraint_disc}
		&\bm{\mathsf{u}}_{N\!-\!1}[k]\in\calU, \quad\; \forall k=0,1,\dots,N-1,\\
		\label{seq:H terminal constraint_disc}
		& \bm{\mathsf{x}}_{N}[\kappa]\in\calF, \qquad \text{for some } \kappa\in\{0,1,\dots,N\},
	\end{align}
\end{subequations}
where $ f_{d}: \bbR^{n}\times\bbR^{m}\rightarrow\bbR^{n} $ denotes the discretized dynamics of system~\eqref{eq:dynamics}. A good approximation of the discretized dynamics can be obtained via numerical integration algorithms, e.g., a Runge-Kutta method. The input trajectory $ \bm{\mathsf{u}}_{N-1} $ and the times $ k $ and $ \kappa $ that solve~\eqref{eq:H_disc} for initial value $ x_{0} $ are denoted by $ \bm{\mathsf{u}}_{N-1,x_{0}}^{\ast} $, $ k_{x_{0}}^{\ast} $ and $ \kappa_{x_{0}}^{\ast} $. Note that due to the incremental definition of the discrete-time dynamics in~\eqref{seq:H dynamics_disc}, $ \bm{\mathsf{u}}_{N-1} $ has one entry less than $ \bm{\mathsf{x}}_{N} $.
By increasing~$ N $, \eqref{eq:H_disc} becomes an arbitrarily close approximation of~\eqref{eq:finite horizon construction H}. 

\subsection{Bypassing the Nested Optimization Problem} 

Max-min-problem~\eqref{eq:finite horizon construction H} allows for an efficient implementation bypassing the nested optimization problem. The essence of this approach is to approximate the inner minimization using a $ p $-norm. 

Let us first consider the discrete-time approximation~\eqref{eq:H_disc}. By drawing a factor $ (-1) $ out from the nested optimization in~\eqref{seq:H max min_disc}, the $ \max $- and $ \min $-operators are reversed and we obtain
\begin{align}
	\label{eq:avoiding-nested-optimization-02}
	H_{T}(x_{0}) := -\min_{\substack{\bm{\mathsf{u}}_{N\!-\!1} \\ \text{s.t. \eqref{seq:H initial condition_disc}-\eqref{seq:H terminal constraint_disc}}}} \max_{k=0,1,\dots,N} -(h(\bm{\mathsf{x}}_{N}[k]) - \gamma k \Delta t).
\end{align}
In the following, we first determine $ \bm{\mathsf{u}}_{N\!-\!1,x_{0}}^{\ast} $ and $ k^{\ast}_{x_{0}} $ before actually computing $ H_{T}(x_{0}) $. To this end, we observe that for some strictly positive function $ f $ it holds 
\begin{align*}
	\argmax_{x} -f(x) = \argmax_{x}\frac{1}{f(x)}
\end{align*}
since $ -f(x_{1}) \!\!>\!\! -f(x_{2}) \Leftrightarrow f(x_{1}) \!\!<\!\! f(x_{2}) \Leftrightarrow \frac{1}{f(x_{2})} \!\!<\!\! \frac{1}{f(x_{1})} $ (replacing the negation with the multiplicative inverse is order-preserving). Furthermore, it is $ \arg\max\frac{1}{f(x)} = \arg\max\frac{1}{f(x)+c} $ for any positive constant $ c\in\bbR_{>0} $. With this, we obtain~$ \bm{\mathsf{u}}_{N-1,x_{0}}^{\ast} $ and~$ k^{\ast}_{x_{0}} $ from~\eqref{eq:avoiding-nested-optimization-02} as
\begin{align}
	\label{eq:avoiding-nested-optimization-03}
	\bm{\mathsf{u}}_{N\!-\!1,x_{0}}^{\ast}, k^{\ast}_{x_{0}} \leftarrow \!\! \min_{\substack{\bm{\mathsf{u}}_{N\!-\!1} \\ \text{s.t. \eqref{seq:H initial condition_disc}-\eqref{seq:H terminal constraint_disc}}}} \max_{k=0,1,\dots,N} \frac{1}{h(\bm{\mathsf{x}}_{N}[k]) \!-\! \gamma k \Delta t \!+ \!\tilde{h}} \!
\end{align}
where $ \tilde{h} \!>\! \max\{0,\min_{x\in\calD} h(x) \!+\! \gamma T) \} $ is some constant that ensures the strict positivity of the denominator. At last for bypassing the nested optimization, we approximate the maximization by a $ p $-norm with a sufficiently high $ p $. Therefore, we note that for any (finite-dimensional) vector $ x=[x_{1},\dots,x_{n}]\in\bbR^{n}_{>0} $ it holds that $ ||x||_{p} \approx \max_{k} x_{k} $ with $ p \gg 0 $ where $ ||x||_{p}:=\sqrt[\leftroot{2}\uproot{2}p]{\sum_{k=1}^{n}x_{k}^{p}} $. Thus, for a sufficiently large $ p $, we obtain
\begin{align}
	\label{eq:avoiding-nested-optimization-04}
	\bm{\mathsf{u}}_{N\!-\!1,x_{0}}^{\ast} \approx  \argmin_{\substack{\bm{\mathsf{u}}_{N\!-\!1} \\ \text{s.t. \eqref{seq:H initial condition_disc}-\eqref{seq:H terminal constraint_disc}}}} \left|\left| 
	\begin{bmatrix}
		\frac{1}{h(\bm{\mathsf{x}}_{N}[0]) - 0\,\gamma\Delta t + \tilde{h}} \\
		\vdots \\
		\frac{1}{h(\bm{\mathsf{x}}_{N}[k]) - k\,\gamma\Delta t + \tilde{h}} \\
		\vdots \\
		\frac{1}{h(\bm{\mathsf{x}}_{N}[N]) - N\,\gamma\Delta t + \tilde{h}}
	\end{bmatrix}
	\right|\right|_{p}.
\end{align}
From this, we can approximate $ \bm{\mathsf{x}}_{N,x_{0}}^{\ast}[k\!+\!1] = f_{d}(\bm{\mathsf{x}}_{N,x_{0}}^{\ast}[k],\bm{\mathsf{u}}_{N-1,x_{0}}^{\ast}[k]) $, $ k=0,1,\dots,N\!-\!1 $, with initial condition $ \bm{\mathsf{x}}_{N,x_{0}}^{\ast}[0] := x_{0} $, and 
\begin{align*}
	k^{\ast}_{x_{0}} = \argmax_{k=0,1,\dots,N} \frac{1}{h(\bm{\mathsf{x}}_{N,x_{0}}^{\ast}[k]) - \gamma k \Delta t + \tilde{h}},
\end{align*}
or equivalently
\begin{align}
	\label{eq:avoiding-nested-optimization-05}
	k^{\ast}_{x_{0}} = \argmin_{k=0,1,\dots,N} h(\bm{\mathsf{x}}_{N,x_{0}}^{\ast}[k]) - \gamma k \Delta t.
\end{align}
From this, we finally obtain
\begin{align}
	\label{eq:avoiding-nested-optimization-06}
	H_{T}(x_{0}) &= h(\bm{\mathsf{x}}_{N,x_{0}}^{\ast}[k^{\ast}_{x_{0}}]) - \gamma k^{\ast}_{x_{0}} \Delta t.
\end{align}
The algorithm is summarized in Algorithm~\ref{algo:finite horizon construction H}.

\begin{algorithm}[t]
	\SetAlgoLined
	\caption{Computation of $ H_{T}(x_{0}) $}
	\Parameters{$ h $, $ f_{d} $, $ T $, $ N $, $ \calF $, $ \gamma $}
	\Input{$ x_{0} $}
	\Output{$ H_{T}(x_{0}) $}
	$ \Delta t  \leftarrow  T/N $\;
	$ \bm{\mathsf{u}}_{N-1,x_{0}}^{\ast}  \leftarrow  \text{solve \eqref{eq:avoiding-nested-optimization-04}}$\;
	$ \bm{\mathsf{x}}_{N,x_{0}}^{\ast}[0] \leftarrow x_{0} $\;
	$ \bm{\mathsf{x}}_{N,x_{0}}^{\ast}[k\!\!+\!\!1] \leftarrow f_{d}(\bm{\mathsf{x}}_{N,x_{0}}^{\ast}[k],\bm{\mathsf{u}}_{N\!-\!1,x_{0}}^{\ast}[k]) $ for $ k\!=\!0,1,\dots,N\!-\!1 $\;
	$H_{T}(x_{0})  \leftarrow  \min_{k=0,1,\dots,N} h(\bm{\mathsf{x}}_{N,x_{0}}^{\ast}[k]) - \gamma k \Delta t $\;
	\label{algo:finite horizon construction H}
\end{algorithm}

\subsection{Terminal Constraint} 

Optimization problem~\eqref{eq:H_disc} is a mixed integer problem due to~\eqref{seq:H terminal constraint_disc}. When $ \calF $ however is readily given as a forward control invariant set, a mixed integer problem can be avoided and~\eqref{eq:H_disc} becomes 
\begin{align}
	\label{eq:H_terminal_disc}
	\begin{split}
		%\label{seq:H_terminal max min_disc}
		H_{T}(x_{0}) &:= \max_{\bm{\mathsf{u}}_{N\!-\!1}} \min_{k=0,1,\dots,N} h(\bm{\mathsf{x}}_{N}[k]) - \gamma k \Delta t \\
		%\label{seq:H_terminal constraints}
		\text{s.t.}\;\; &\text{\eqref{seq:H initial condition_disc}-\eqref{seq:H input constraint_disc} hold,}\\
		%\label{seq:H_terminal terminal constraint_disc}
		& \bm{\mathsf{x}}_{N}[N]\in\calF.
	\end{split}
\end{align}
If $ \calF $ is forward control invariant, Assumption~\ref{ass:setF} is still satisfied and Theorem~\ref{thm:predictive CBF} applies.

\subsection{Computation of the Saturated CBF $ \bar{H}_{T} $} 

The computation of $ \widebar{H}_{T} $ reduces the effort for computing a CBF in the Dini sense, as $ \widebar{H}_{T}(x_{0}) = \delta - \gamma T $ for $ x_{0}\in\calF $. Thus, $ \widebar{H}_{T} $ is only required to be explicitly computed on $ \calD\setminus\calF $.

%%%%%%%%%%%%%%%%%%%%%%%%%%%%%%%%%%%%%%%%%%%%%%%%%%%%%%%%%%%%%%%%%%%%%%%%%%%%%%%%
% SIMULATIONS

\section{Numerical Examples}
\label{sec:simulations}

In the following, we apply our CBF synthesis method to the design of safety filters for various dynamic systems in the presence of both static and time-varying constraints. In particular, we consider 
\begin{enumerate}[label=\Alph*.]
	\item single and double integrators as examples for first and second order systems; 
	\item the bicycle model with minimum velocity $ >0 $ as an example for a system that is not locally controllable;
	\item the unicycle as an example for a system with a non-holonomic constraint and state-dependent relative degree.
\end{enumerate}
For each of the systems, we compute the function $ H_{T} $ based on~\eqref{eq:finite horizon construction H} for the static constraint over a grid and approximate all further values by linear interpolation. The approximated function coincides in each of the grid points with $ H_{T} $, while all further points are close approximations. Also other methods for fitting a function into the set of computed values can be applied including the training of a neural network. The subsequent simulation examples however indicate that also elementary approximation methods yield promising results. An implementation of our CBF synthesis method in Python using Casadi~\cite{Andersson2019} is provided on Github\footnote{Implementation and examples on Github: \url{https://github.com/KTH-DHSG/Predictive-CBF-Synthesis-Toolbox.git}}. It allows for the parallelized computation of $ H_{T} $. Videos to the examples can be found on Youtube\footnote{Videos to the examples: \url{https://www.youtube.com/watch?v=8inhub7IhFY}}.

\begin{table*}[t]
	\centering
	\begin{tabular}{l!{\vrule}c!{\vrule}cccc!{\vrule}cc!{\vrule}ccccc}
		\toprule
		\multicolumn{1}{c}{} & \multicolumn{1}{c}{\textit{constr.}} &  \multicolumn{4}{c}{\textit{CBF params.}} & \multicolumn{2}{c}{\textit{class~$ \calK_{e} $ fcn.}} & \multicolumn{5}{c}{\textit{numerical computation}} \\
		& \textbf{$ r $} & \textbf{$ \gamma $} & \textbf{$ \delta $} & \textbf{$\overline{T}$} & \textbf{$ T $} &  \textbf{$ c $} & \textbf{$ c_{\alpha} $} & \textbf{$ N $} & \textbf{domain} & \textbf{discretization} &   \textbf{\#points} & \textbf{comp. time} \\
		\midrule
		single integrator & 9 & 2 & 1 & 0 & 10 & 2 & 0.2 & 25 & $ [-10,10]^{2} $ & $ 41,41 $ & 1681 & $ 0\!:\!01\!:\!24 $\\
		single i. ($ v_{x,\text{min}} > 0 $) & 9 & 2 & 1 & 0 & 10 & 2 & 0.2 & 25 & $ [-10,10]^{2} $ & $ 41,41 $ & 1681 & $ 0\!:\!01\!:\!11 $ \\
		double integrator & 9 & 1 & 5 & 3 & 12 & 0.5 & 0.2 & 30 & $ [-14,14]^{2} \times [-2.5,2.5]^{2} $ & $ 29,29,15,15 $ & 189 225 & $ 2\!:\!49\!:\!46 $ \\
		bicycle (less agile) & 6 & 2 & 9 & 4.4 & 12 & 1 & 0.1 & 30 & $ [-15,15]^{2} \times [-\pi,\pi] $ & $ 61,61,41 $ & 152 561 & $ 5\!:\!01\!:\!42 $ \\
		bicycle (more agile) & 6 & 2 & 4 & 1.8 & 10 & 1 & 0.1 & 30 & $ [-10,10]^{2} \times [-\pi,\pi] $ & $ 41,41,41 $ & 68 921 & $ 2\!:\!33\!:\!33 $ \\
		unicycle & 6 & 2 & 4 & 1.8 & 10 & 0.5 & 0.1 & 30 & $ [-10,10]^{2} \times [-\pi,\pi] $ & $ 41,41,41 $ & 68 921 & $ 1\!:\!38\!:\!12 $ \\
		\bottomrule
	\end{tabular}
	\caption{Constraint specs, parameters and computation times [h:mm:ss] for the CBFs of each considered dynamic system.}
	\label{tab:computation time}
	\vspace{-\baselineskip}
\end{table*}

The constraint function under consideration is 
\begin{align*}
	h(x) = \sqrt{(x-x_{c})^{2} + (y-y_{c})^{2}} - r
\end{align*}
describing a circular obstacle with center $ (x_{c},y_{c}) $ and radius~$ r $; the square root ensures that $ h $ scales linearly with the distance. The extended class~$ \calK_{e} $ functions are chosen as convex functions of the form $ \alpha(\zeta) \coloneq \left\{\begin{smallmatrix*}[l]
	c\zeta & \text{if } \zeta \geq 0 \\
	2\gamma \, \left(\text{sig}\left(\tfrac{c\zeta}{4}\right) - 0.5\right) & \text{if } \zeta<0
\end{smallmatrix*}\right. $ according to Theorem~\ref{thm:predictive CBF} with~$ c>0 $. To vary the constraint function $ h $ and the synthesized CBF $ H_{T} $ over time, we add the periodic time-varying function
\begin{align*}
%	\label{eq:lambda}
	\bm{\lambda}(t) = -r_{\text{max}} \, \left|\sin\left(\tfrac{\pi t}{\tau_{p}} - \sigma\right)\right| + r
\end{align*}
where $ r_{\text{max}} \leq r $ denotes the maximum radius, $ \tau_{p} $ is the period and $ \sigma $ some shift. Its parameters are chosen such that~\eqref{eq:lambda condition simplified} holds and thereby Proposition~\ref{prop:lambda shiftable} applies. Thus, $ H_{\bm{\lambda}(\cdot),T}(t,x) \coloneq H_{T}(x) + \bm{\lambda}(t) $ is a CBF. The control task can be now stated as follows: track a straight line while avoiding all possibly time-varying obstacles. For tracking, we employ a feedback controller that generates a baseline input $ u_{\text{baseline}} $. For obstacle avoidance, we use a standard safety filter based on the time-varying CBF $ H_{\bm{\lambda}(\cdot),T} $ defined~as
\begin{align*}
	u_{\text{safe}}(t) &= \argmin_{u\in\calU} (u-u_{\text{baseline}}(t))^{T} P (u-u_{\text{baseline}}(t)) \\
	\text{s.t. } &dH_{T}(\bm{x}(t);f(\bm{x}(t),\bm{u}(t))) + \dot{\bm{\lambda}}(t)\\
	&\hspace{1.5cm} \geq-\alpha(H_{T}(\bm{x}(t))+\bm{\lambda}(t)) + c_{\alpha}
\end{align*}
where $ P $ is some positive definite matrix, and $ c_{\alpha}\geq0 $ some constant that can be set positive to add robustness with view to the discretization. 
An overview over all parameters and computation times of the various CBFs can be found in Table~\ref{tab:computation time}. The CBFs have been computed on a 12th Gen Intel Core i9-12900K with 64GB RAM.

\subsection{Single and Double Integrator}

\begin{figure}[t]
	\centering
	\begin{subfigure}[b]{0.5\linewidth}
		\centering
		\includegraphics[width=\linewidth]{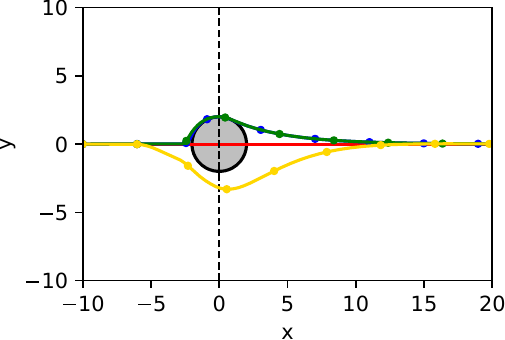}
		\caption{}
		\label{sub_fig:sc_traj_static}
	\end{subfigure}
	\hfill
	\begin{subfigure}[b]{0.4\linewidth}
		\centering
		\includegraphics[width=\linewidth]{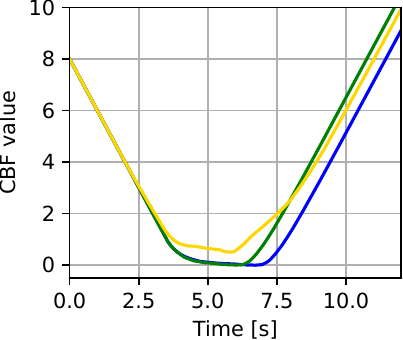}
		\caption{}
		\label{sub_fig:sc_cbf_static}
	\end{subfigure}
	\linebreak
	\begin{subfigure}[b]{0.5\linewidth}
		\centering
		\includegraphics[width=\linewidth]{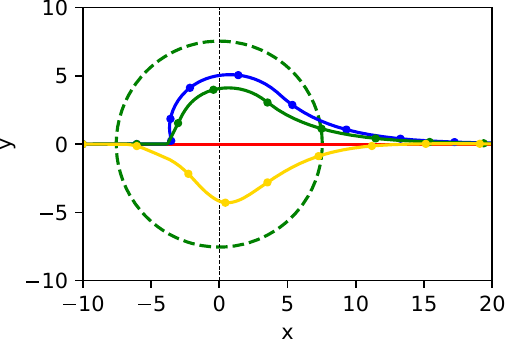}
		\caption{}
		\label{sub_fig:sc_traj_tv}
	\end{subfigure}
	\hfill
	\begin{subfigure}[b]{0.4\linewidth}
		\centering
		\includegraphics[width=\linewidth]{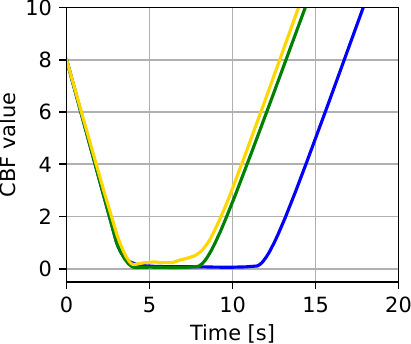}
		\caption{}
		\label{sub_fig:sc_cbf_tv}
	\end{subfigure}
	\linebreak
	\begin{subfigure}[b]{0.45\columnwidth}
		\centering
		\includegraphics[width=\linewidth]{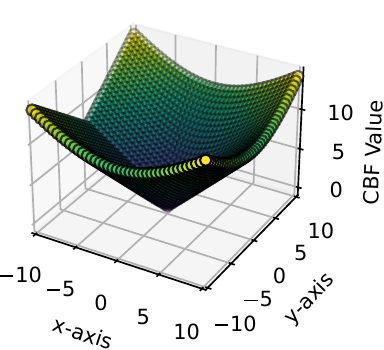}
		\caption{CBF for the single integrator with $ \dot{x}\in[1,2] $. Points mark explicitly computed values.}
		\label{sub_fig:s1_forward_cbf}
	\end{subfigure}
	\hfill
	\begin{subfigure}[b]{0.47\columnwidth}
		\centering
		\includegraphics[width=\linewidth]{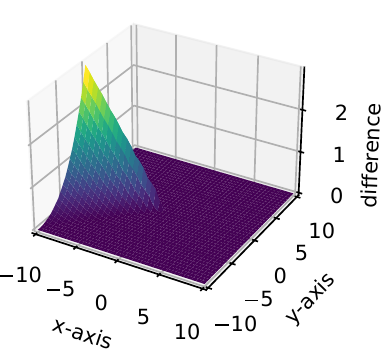}
		\caption{Difference between constraint function $ h $ and the CBF depicted in Fig.~\ref{sub_fig:s1_forward_cbf}.}
		\label{sub_fig:s1_forward_difference}
	\end{subfigure}
	\caption{Simulation results for single and double integrators: trajectories and the corresponding CBF values are depicted in (a), (b) for static and in (c), (d) for time-varying constraints (maximum expansion is indicated by the dotted circle). The single integrator with $ \calU=[-2,2]^{2} $ is marked blue, the one with $ \calU=[1,2]\times[-2,2] $ in green, and the double integrator in yellow. (e), (f) depict a CBF and compare it to~$ h $.}
	\label{fig:sc}
	\vspace{-\baselineskip}
\end{figure}

First, we consider single and double integrators in a plane with dynamics $ \dot{\mathbf{x}} = u $ and $ \ddot{\mathbf{x}} = u $ where $ \mathbf{x} = [x,y]^{T} $. We consider two single integrators where the first (marked blue in Figure~\ref{fig:sc}) has input constraints $ \calU=[-2,2]^{2} $ and the second (green) $ \calU = [1,2]\times[-2,2] $. While the first single integrator can stop or move backwards, the second always moves into the positive $ x $-direction. For the double integrator (yellow), the input constraint is $ \calU=[-1,1]^{2} $, and we limit its velocity to $ \dot{\mathbf{x}}\in[-2,2]^{2} $ using the modified constraint function $ \tilde{h}(\mathbf{x},\dot{\mathbf{x}}) \coloneq \eta \min\{\frac{1}{\eta}\,h(\mathbf{x}), 2+\dot{x}, 2+\dot{y}, 2-\dot{x}, 2-\dot{y}\} $ with some positive constant $ \eta\in\bbR_{>0} $. The simulation results are depicted in Figure~\ref{fig:sc} for static (a,b) and time-varying constraints (c,d). While in the static case, both single integrators have an almost identical trajectory, in the time-varying case they become distinct as the first single integrator uses its capability to move backwards. The positive CBF values in Figures~\ref{sub_fig:sc_cbf_static} and~\ref{sub_fig:sc_cbf_tv} indicate constraint satisfaction. 

Even in the case of single integrators, the CBF synthesis can be a non-trivial task. While the constraint function $ h $ is directly a CBF for the unconstrained single integrator, the same is not necessarily true in the presence of input constraints. This can be clearly seen in the case of the second single integrator for which $ \dot{x} \in [1,2] $; its numerically computed CBF is depicted in Figure~\ref{sub_fig:s1_forward_cbf}. It can be seen that, in contrast to $ h $, the CBF is nonsmooth along the negative $ x $-axis. The difference between $ h $ and the CBF is depicted in Figure~\ref{sub_fig:s1_forward_difference}.

\subsection{Kinematic Bicycle Model}

\begin{figure}[!t]
	\centering
	\begin{subfigure}[b]{0.9\columnwidth}
		\centering
		\includegraphics[width=\linewidth]{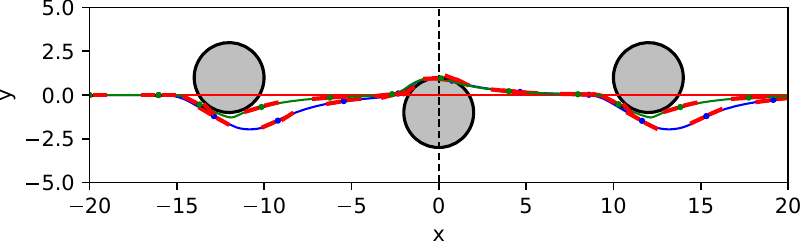}
		\caption{Trajectories of the kinematic bicycle model for static constraints. A marker is depicted every~2s.}
		\label{sub_fig:b1_traj}
	\end{subfigure}
	\linebreak
	\begin{subfigure}[b]{0.9\columnwidth}
		\centering
		\includegraphics[width=\linewidth]{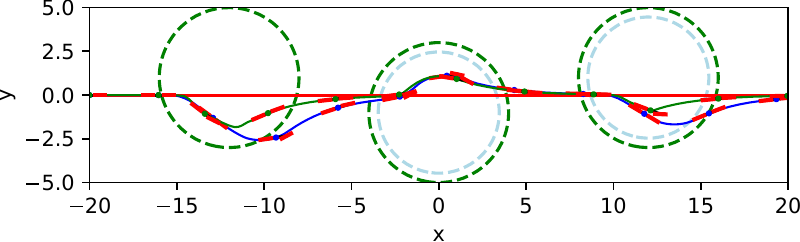}
		\caption{Trajectories of the kinematic bicycle model for time-varying constraints. Green circles mark the maximum and light-blue ones the initial expansion of the obstacles. A marker is depicted every~2s.}
		\label{sub_fig:b1_traj_tv}
	\end{subfigure}
	\linebreak
	\begin{subfigure}[b]{0.39\columnwidth}
		\centering
		\includegraphics[width=\linewidth]{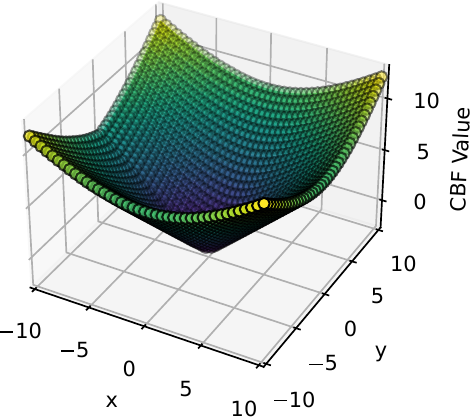}
		\caption{CBF at $ \psi = 0 $ for $ \zeta_{\text{max}} =  20 \pi/180 $ and one circular obstacle. Points mark~the explicitly computed values.}
		\label{sub_fig:b1_2p8_cbf}
	\end{subfigure}
	\hfill
	\begin{subfigure}[b]{0.47\columnwidth}
		\centering
		\includegraphics[width=\linewidth]{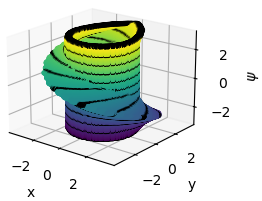}
		\caption{Zero super-level sets in the $ x $-$ y $-plane in dependence of orientation $ \psi $ for $ \zeta_{\text{max}} =  20 \pi/180 $.}
		\label{sub_fig:b1_2p8_cbf_zerosuperlevelset}
	\end{subfigure}
	\linebreak
	\begin{subfigure}[b]{0.8\columnwidth}
		\centering
		\includegraphics[width=\linewidth]{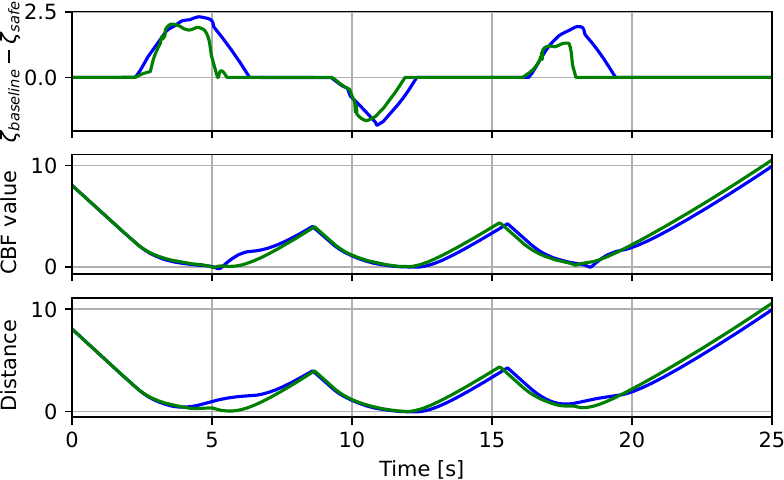}
		\caption{Correction of steering angle, CBF values and distance to obstacle over time (time-varying constraints).}
		\label{sub_fig:b1_cbf_tv}
	\end{subfigure}
	\linebreak
	\begin{subfigure}[b]{0.9\columnwidth}
		\centering
		\includegraphics[width=\linewidth]{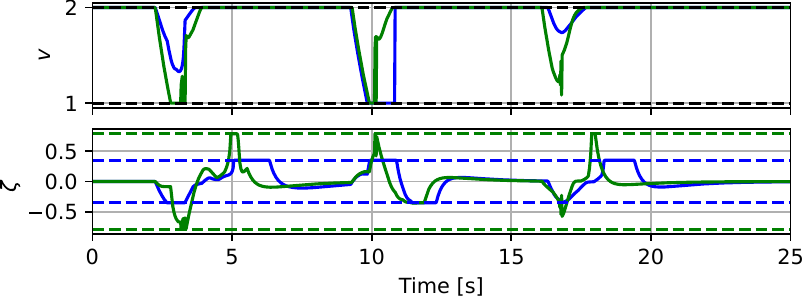}
		\caption{Input trajectories (time-varying constraints).}
		\label{sub_fig:b1_inputs_tv}
	\end{subfigure}
	\caption{Simulation results for the kinematic bicycle model: the bicycle with $ \zeta_{\text{max}} =  20 \pi/180 $ is indicated in blue, the one with $ \zeta_{\text{max}} =  45\pi/180  $ in green.}
	\label{fig:b1}
	\vspace{-\baselineskip}
\end{figure}

Let us reconsider the kinematic bicycle model from Example~\ref{exmp:bicycle1} with inputs $ u=[v,\zeta]^{T} $ and  input constraints $ \calU = [1,2] \times [-\zeta_{\text{max}},\zeta_{\text{max}}] $ where $ \zeta_{\text{max}} $ specifies the limitations of the steering angle. In Figure~\ref{fig:b1}, we consider bicycles with two different input constraints: the first is less agile with $ \zeta_{\text{max}}=\frac{20}{180}\pi $ (blue), while the second is more agile with $ \zeta_{\text{max}}=\frac{40}{180}\pi $ (green). As it can be seen from Figures~\ref{sub_fig:b1_traj} and~\ref{sub_fig:b1_traj_tv}, the more agile bicycle stays closer to the obstacles than the less agile one. Each of the obstacles is encoded via a separate CBF that can be varied in time independently of the others. The sinusoidal signal varying the size of each of the obstacles is shifted by a third period; initial and maximum expansions are indicated by green and light-blue dashed circles in Figure~\ref{sub_fig:b1_traj_tv}. For the case with time-varying obstacles, the difference between the safe and the baseline input, the CBF values and the vehicles' distances to the obstacles are depicted in Figure~\ref{sub_fig:b1_cbf_tv}. Figure~\ref{sub_fig:b1_inputs_tv} shows the corresponding control inputs. It indicates that the time-varying constraints could be satisfied with finite inputs that stay within the input constraints. Because $ v>0 $, the CBF of both bicycles is nonsmooth as well as in the previous example, which is shown in Figure~\ref{sub_fig:b1_2p8_cbf} for $ \zeta_{\text{max}}=\frac{20}{180}\pi $ and a fixed orientation. As it can be seen from Figure~\ref{sub_fig:b1_2p8_cbf_zerosuperlevelset}, the zero super-level sets are highly dependent on the vehicle's orientation. 

\subsection{Kinematic Unicycle Model}

\begin{figure}[!t]
	\centering
	\begin{subfigure}[b]{0.9\columnwidth}
		\centering
		\includegraphics[width=\linewidth]{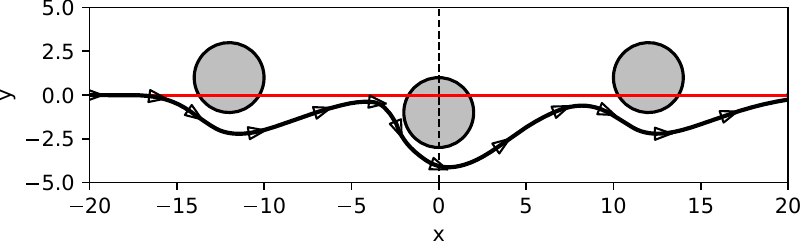}
		\caption{Trajectories of the kinematic unicycle model for static constraints. A marker is depicted every~2s.}
		\label{sub_fig:u1_traj}
	\end{subfigure}
	\linebreak
	\begin{subfigure}[b]{0.9\columnwidth}
		\centering
		\includegraphics[width=\linewidth]{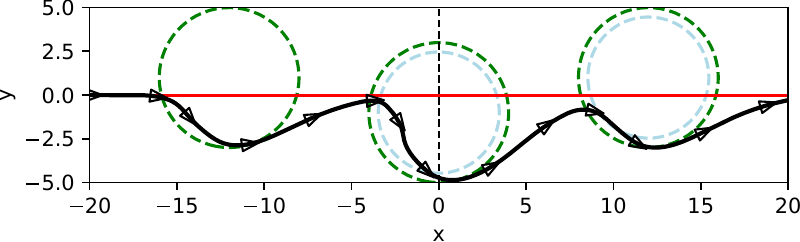}
		\caption{Trajectories of the kinematic unicycle model for time-varying constraints. Green circles mark the maximum and light-blue ones the initial expansion of the obstacles. A marker is depicted every~2s.}
		\label{sub_fig:u1_traj_tv}
	\end{subfigure}
	\linebreak
	\begin{subfigure}[b]{0.8\columnwidth}
		\centering
		\includegraphics[width=\linewidth]{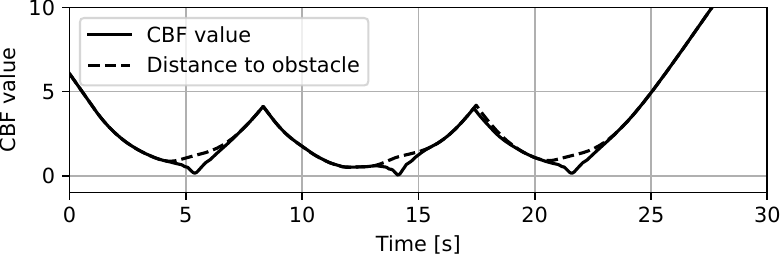}
		\caption{CBF values and distance to obstacle over time (time-varying constraints).}
		\label{sub_fig:u1_cbf_dist_tv}
	\end{subfigure}
	\caption{Simulation results for the kinematic unicycle model.}
	\label{fig:u1}
	\vspace{-\baselineskip}
\end{figure}

As last example, we consider the unicycle dynamics
\begin{align*}
	\dot{x} = v\cos(\psi), \quad 	\dot{y} = v\sin(\psi) , \quad	\dot{\psi} = \omega 
\end{align*}
with input vector $ u = [v,\omega]^{T} $ and input constraint $ \calU=[-0.9,0.9]\times[1,2] $. The dynamic properties of the kinematic unicycle are of particular interest, as it firstly involves a nonholonomic constraint, and secondly its relative degree is state-dependent. As such, the relative degree with respect to state~$ x $ is one for all $ \psi \neq \pi/2 $. Yet, given an orientation of $ \psi=\pi/2 $, one obtains $ \dot{x} = 0 $ and the system becomes a second-order system with respect to~$ x $. An analogous observation holds for state $ y $ and orientation $ \psi = 0 $. As our approach, however, is indifferent to the order of the system, the CBF synthesis method stays the same as for the bicycle dynamics in the previous example. Hence, no high-order CBF concepts need to be employed. Simulations in scenarios analogous to those for the kinematic bicycle model have been conducted; the results are shown in Figure~\ref{fig:u1}. As indicated by Figure~\ref{sub_fig:u1_cbf_dist_tv}, constraint satisfaction is also ensured for the unicycle despite the more challenging dynamic properties. We point out that the values of the CBF and the distance to the obstacle are not trivially correlated due to the dynamic properties of the system. 

%%%%%%%%%%%%%%%%%%%%%%%%%%%%%%%%%%%%%%%%%%%%%%%%%%%%%%%%%%%%%%%%%%%%%%%%%%%%%%%%
% CONCLUSION

\section{Conclusion}
\label{sec:conclusion}

In this work, we presented a systematic method for synthesizing CBFs that encode predictive information. We showed how this information can be advantageously used to account for changes in the constraint specifications and to derive time-varying CBFs. In particular, we presented three synthesis methods that allow to specify the time-varying capabilities of the CBF in terms of a design parameter. The theoretical analysis of the synthesized CBFs was complemented by a detailed discussion of its properties and practical implementation remarks. The proposed method was applied to multiple dynamic systems to demonstrate its applicability. This work is accompanied by a python implementation of the synthesis method that allows for parallelization. 
The challenge that CBF synthesis methods generally do not scale well with dimensionality remains and has not been addressed in this work. In our follow-up work, we show how the synthesis method presented here can be advantageously applied to equivariant systems in order to reduce the time complexity of the CBF synthesis. Furthermore, compositional controller design approaches may be of interest in this direction.

%%%%%%%%%%%%%%%%%%%%%%%%%%%%%%%%%%%%%%%%%%%%%%%%%%%%%%%%%%%%%%%%%%%%%%%%%%%%%%%%
%\hspace{-0.0cm}
%\section*{References}
%
%\vspace*{-1.5em} 

\bibliographystyle{IEEEtran}
\bibliography{/Users/wiltz/CloudStation/JabBib/Research/000_MyLibrary}

%%%%%%%%%%%%%%%%%%%%%%%%%%%%%%%%%%%%%%%%%%%%%%%%%%%%%%%%%%%%%%%%%%%%%%%%%%%%%%%%

% APPENDIX

\section*{Appendix}

\begin{proof}[Proof of Proposition~\ref{prop:setF eps}]
	Let us construct in the following for each $ x_{0}\in\partial\calF $ an input trajectory $ \bm{u}_{x_{0}}\in\bm{\calU} $ and its corresponding state trajectory $ \bm{\varphi}(t;x_{0},\bm{u}_{x_{0}}) $, which we consider over a time horizon $ t\in[0,t_{f,x_{0}}] $ where $ t_{f,x_{0}} $ is to be specified later. As before, we consider $ \calV \coloneq \bigcup_{x_{0}\in\partial\calF} \{ x \, | \, x = \bm{\varphi}(\tau;x_{0},\bm{u}_{x_{0}}), \; \tau\in[0,t_{f,x_{0}}] \} \cup \calF $ for which clearly $ \calF\subseteq\calV $. It remains to show that for each condition, $ \bm{u}_{x_{0}} $ and $ t_{f,x_{0}} $ can be chosen such that $ \calV $ is forward control invariant and $ h(x)\geq\delta $ for all $ x\in\calV $.
	
	Let condition~\ref{cond:prop:set F eps locally-locally controllable} hold. Consider a state $ x_{f,x_{0}}\in\calF \cap B_{\delta_{\varepsilon}}(x_{0}) $ in the neighborhood of an arbitrary $ x_{0}\in\partial\calF $. As system~\eqref{eq:dynamics} is locally-locally controllable in any $ x_{0}\in\partial\calF $, there exists by definition a $ \bm{u}_{x_{0}} $ and $ t_{f,x_{0}} $ such that $ \bm{\varphi}(t_{f,x_{0}};x_{0},\bm{u}_{x_{0}})= x_{f,x_{0}} $ and $ \bm{\varphi}(t;x_{0},\bm{u}_{x_{0}})\in B_{\varepsilon}(x_{0}) $ for all $ t\in[0,t_{f,x_{0}}] $. Then by Proposition~\ref{prop:setF}, the above defined $ \calV $ is forward control invariant. Moreover, as for any $ x_{0}\in\calF $ it holds $ \bm{\varphi}(t;x_{0},\bm{u}_{x_{0}})\in \calF\oplus B_{\varepsilon}(x_{0}) $ for all $ t\in[0,t_{f,x_{0}}] $, we have $ \calV\subseteq\calF \oplus B_{\varepsilon}(0) $. From this, we conclude that $ \calV\subseteq\calF \oplus B_{\varepsilon}(0) \subseteq \{x \,|\, h(x)\geq\delta\} $ where the last subset relation holds by~\cite[(3.1.12)]{Schneider2014}. Hence, Assumption~\ref{ass:setF} holds.
	
	Let now condition~\ref{cond:prop:set F eps ball around x0} hold. For each $ x_{0}\in\partial\calF $, choose $ \bm{u}_{x_{0}} $ as suggested by the condition and let $ t_{f,x_{0}} \rightarrow \infty $. Then, $ \calV $ is clearly forward control invariant and $ \calV \subseteq \calF \oplus B_{\varepsilon}(0) \subseteq \{x \,|\, h(x)\geq\delta\} $ as previously. Thus, Assumption~\ref{ass:setF} holds.
	
	Let at last condition~\ref{cond:prop:set F eps nbhd of Fhat} hold. Choose $ u_{x_{0}} $ and $ t_{f,x_{0}} $ as suggested by the condition, and it follows from analogous arguments as in the previous case that Assumption~\ref{ass:setF} holds.
\end{proof}

\begin{proof}[Proof of Proposition~\ref{prop:finite horizon CBF construction impact time horizon}]
	Let us consider a trajectory $ \bm{\varphi}(t;x_{0},\bm{u}_{x_{0},T_{1}}^{\ast}) $ defined for $ t\in[0,T_{1}] $ starting in an arbitrary point $ x_{0}\in\calD $. The input trajectory and those times that solve~\eqref{eq:finite horizon construction H} for time horizon $ T_{1} $ are denoted by $ \bm{u}_{x_{0},T_{1}}^{\ast} $, $ t_{x_{0},T_{1}}^{\ast} $ and $ \vartheta_{x_{0},T_{1}}^{\ast} $, respectively. Similarly to before, we define an extended input trajectory as $ \bm{u}_{e,x_{0},T_{1}}^{\ast}(t) \coloneq \left\{ \begin{smallmatrix*}[l]
		\bm{u}^{\ast}_{x_{0},T_{1}}(t) & \text{if } t\in[0,\vartheta^{\ast}_{x_{0},T_{1}}] \\
		\bm{u}_{e,x_{0},T_{1}}(t) & \text{if } t>\vartheta^{\ast}_{x_{0},T_{1}}
	\end{smallmatrix*} \right. $
	%	\begin{align*}
		%		%		\label{eq:prop:finite horizon CBF construction impact time horizon aux 0}
		%		\bm{u}_{e,x_{0},T_{1}}^{\ast}(t) \coloneq 
		%		\begin{cases}
			%			\bm{u}^{\ast}_{x_{0},T_{1}}(t) &\text{if } t\in[0,\vartheta^{\ast}_{x_{0},T_{1}}] \\
			%			\bm{u}_{e,x_{0},T_{1}}(t) &\text{if } t>\vartheta^{\ast}_{x_{0},T_{1}}
			%		\end{cases}
		%	\end{align*}
	where $ \bm{u}_{e,x_{0},T_{1}}\in\bm{\calU}_{(\vartheta^{\ast}_{x_{0}},\infty)} $ such that $ \bm{\varphi}(t;x_{0},\bm{u}_{e,x_{0},T_{1}}^{\ast})\in\calV $ for all $ t>\vartheta^{\ast}_{x_{0},T_{1}} $. Since $ \calV $ is forward control invariant, such a trajectory exists. 
	
	Let us first consider those states $ x_{0} $ with $ H_{T_{1}}(x_{0})<\delta - \gamma T_{2} $, where the right-hand side intentionally employs~$ T_{2} $ instead of~$ T_{1} $. Furthermore, note that the right-hand side is strictly positive due to~\eqref{eq:prop:finite horizon CBF construction impact time horizon 0}. We now derive analogously to~\eqref{eq:thm:predictive CBF aux 2} that
	\begin{align}
		&H_{T_{1}}(x_{0}) = h(\bm{\varphi}(t^{\ast}_{x_{0},T_{1}};x_{0},\bm{u}_{x_{0},T_{1}}^{\ast})) - \gamma t_{x_{0},T_{1}}^{\ast} \nonumber\\
		&\quad\leq \delta - \gamma T_{2}  \nonumber\\
		\label{eq:prop:finite horizon CBF construction impact time horizon aux 1}
		&\quad\leq h(\bm{\varphi}(t;x_{0},\bm{u}_{e,x_{0},T_{1}}^{\ast})) - \gamma t \quad \forall t\in[\vartheta^{\ast}_{x_{0},T_{1}},T_{2}].
	\end{align}
	Based on this, we further obtain  
	\begin{subequations}
		\label{eq:prop:finite horizon CBF construction impact time horizon aux 2}
		\begin{align}
			\label{seq:prop:finite horizon CBF construction impact time horizon aux 2.1}
			&H_{T_{1}}(x_{0}) = h(\bm{\varphi}(t^{\ast}_{x_{0},T_{1}};x_{0},\bm{u}_{x_{0},T_{1}}^{\ast})) - \gamma t_{x_{0},T_{1}}^{\ast} \\
			\label{seq:prop:finite horizon CBF construction impact time horizon aux 2.2}
			&\qquad= \min_{t\in[0,T_{1}]} h(\bm{\varphi}(t;x_{0},\bm{u}^{\ast}_{x_{0},T_{1}})) - \gamma t \\
			\label{seq:prop:finite horizon CBF construction impact time horizon aux 2.3}
			&\qquad\stackrel{\eqref{eq:prop:finite horizon CBF construction impact time horizon aux 1}}{=} \min_{t\in[0,T_{2}]} h(\bm{\varphi}(t;x_{0},\bm{u}_{e,x_{0},T_{1}}^{\ast})) - \gamma t \\
			\label{seq:prop:finite horizon CBF construction impact time horizon aux 2.4}
			&\qquad\leq \min_{t\in[0,T_{2}]} h(\bm{\varphi}(t;x_{0},\bm{u}_{x_{0},T_{2}}^{\ast})) - \gamma t = H_{T_{2}}(x_{0})
		\end{align}
	\end{subequations}
	where~\eqref{seq:prop:finite horizon CBF construction impact time horizon aux 2.3} follows analogously to~\eqref{eq:thm:predictive CBF aux 3} as it holds for $ T_{2} $ that $ \vartheta_{x_{0},T_{1}}^{\ast} \leq T_{1} \leq T_{2} $, \eqref{seq:prop:finite horizon CBF construction impact time horizon aux 2.4} follows due to the suboptimality of $ \bm{u}_{e,x_{0},T_{1}}^{\ast} $, and $ \bm{u}_{x_{0},T_{2}}^{\ast} $ denotes the input trajectory that solves~\eqref{eq:finite horizon construction H} for time horizon $ T_{2} $ at state $ x_{0} $. 
	
	At last, we note that for the remaining states $ x_{0} $ with $ H_{T_{1}}(x_{0})> \delta - \gamma T_{2} $ we have that $ H_{T_{2}}(x_{0})>0 $ as
	\begin{subequations}
		\label{eq:prop:finite horizon CBF construction impact time horizon aux 3}
		\begin{align}
			\label{seq:prop:finite horizon CBF construction impact time horizon aux 3.1}
			&H_{T_{2}}(x_{0}) = \min_{t\in[0,T_{2}]} h(\bm{\varphi}(t;x_{0},\bm{u}_{x_{0},T_{2}}^{\ast})) - \gamma t \\
			\label{seq:prop:finite horizon CBF construction impact time horizon aux 3.2}
			&\quad\geq \min_{t\in[0,T_{2}]} h(\bm{\varphi}(t;x_{0},\bm{u}_{e,x_{0},T_{1}}^{\ast})) - \gamma t \\
			\label{seq:prop:finite horizon CBF construction impact time horizon aux 3.3}
			&\quad= \min\!\left\{ \!H_{T_{1}}\!(x_{0}), \!\min_{t\in[T_{1},T_{2}]}\! h(\bm{\varphi}(t;\!x_{0},\!\bm{u}_{e,x_{0},T_{1}}^{\ast})) \!-\! \gamma t  \right\} \\
			\label{seq:prop:finite horizon CBF construction impact time horizon aux 3.4}
			&\quad\geq \delta - \gamma T_{2} \stackrel{\eqref{eq:prop:finite horizon CBF construction impact time horizon 0}}{\geq} 0
		\end{align}
	\end{subequations}
	where~\eqref{seq:prop:finite horizon CBF construction impact time horizon aux 3.2} follows from the suboptimality of $ \bm{u}_{e,x_{0},T_{1}}^{\ast} $, and~\eqref{seq:prop:finite horizon CBF construction impact time horizon aux 3.4} holds as $ \bm{\varphi}(t;x_{0},\bm{u}_{e,x_{0},T_{1}}^{\ast})\in\calV $ for all $ t\in[T_{1},T_{2}] $, $ h(x)>\delta $ for all $ x\in\calV $ and $ \gamma t < \gamma T_{2} $ for all $ t\in[T_{1},T_{2}] $. 
	
	We have now shown that the following holds: $ H_{T_{1}}(x_{0}) \geq 0 $ implies $ H_{T_{2}}(x_{0}) \geq 0 $, and thus $ \calC_{0,T_{1}} \subseteq \calC_{0,T_{2}} $; moreover, as for all $ x_{0}\in\calD $ with $ H_{T_{1}}(x_{0})<\delta - \gamma T_{2} $, and in particular for all $ x_{0}\in\calD $ with $ H_{T_{1}}\leq 0 $, it holds $ H_{T_{1}}(x_{0})\leq H_{T_{2}}(x_{0}) $ as by~\eqref{eq:prop:finite horizon CBF construction impact time horizon aux 2}, we conclude that also $ \calC_{\lambda,T_{1}} \subseteq \calC_{\lambda,T_{2}} $ for all $ \lambda\geq0 $ with $ \calC_{\lambda,T_{1}}, \calC_{\lambda,T_{2}} \subseteq\calD $. 
\end{proof}

%%%%%%%%%%%%%%%%%%%%%%%%%%%%%%%%%%%%%%%%%%%%%%%%%%%%%%%%%%%%%%%%%%%%%%%%%%%%%%%%

\end{document}